\documentstyle[prl,floats,aps,twocolumn,epsf]{revtex}
\tighten

\begin{document}
\draft

\twocolumn[\hsize\textwidth\columnwidth\hsize\csname
@twocolumnfalse\endcsname

\title{Reconstruction of inhomogeneous metric perturbations and
electromagnetic four-potential in Kerr spacetime}
\author{Amos Ori}
\address{Department of Physics, Technion---Israel Institute of Technology,
Haifa, 32000, Israel}
\date{\today}
\maketitle

\begin{abstract}
We present a procedure that allows the construction of the metric
perturbations and electromagnetic four-potential, for gravitational and
electromagnetic perturbations produced by sources in Kerr spacetime. This
may include, for example, the perturbations produced by a point particle or
an extended object moving in orbit around a Kerr black hole. 
The construction is carried out in the frequency domain. Previously,
Chrzanowski derived the vacuum metric perturbations and electromagnetic
four-potential by applying a differential operator to a certain potential $\Psi $. Here we construct $\Psi $ for inhomogeneous perturbations, thereby
allowing the application of Chrzanowski's method. We address this problem in
two stages: First, for vacuum perturbations (i.e. pure gravitational or
electromagnetic waves), we construct the potential from the modes of the
Weyl scalars $\psi _{0}$ or $\varphi _{0}$. Second, for perturbations
produced by sources, we express $\Psi $ in terms of the mode functions of
the source, i.e. the energy-momentum tensor $T_{\alpha \beta }$ or the
electromagnetic current vector $J_{\alpha }$.

\end{abstract}


\vspace{2ex}
]

%**********************************************************************
%**********************************************************************
%**********************************************************************

\section{Introduction}

\label{sec1}

The gravitational perturbations of Kerr Black holes (BHs) are fully
described by the metric perturbation (MP) $h_{\alpha \beta }$. The latter
satisfies the linearized Einstein equation, which is a set of coupled,
linear, partial differential equations. Teukolsky \cite{t0}\cite{t1} showed,
however, that the curvature Weyl scalars $\psi _{0}$ and $\psi _{4}$ each
satisfies a decoupled field equation, the ''master equation''. Furthermore,
this decoupled equation may be separated, leading to ordinary differential
equations for the radial and angular parts. This leads to a great
simplification of the problem of determining the gravitational perturbations.

The problem of electromagnetic perturbations over a Kerr background has a
similar status. The Maxwell equations form a set of coupled, linear, partial
differential equations for the four-potential $A_{\alpha }$ (or for the Maxwell
field $F_{\alpha \beta }$). In this case, too, Teukolsky \cite{t0}\cite{t1}
derived separable, decoupled, equations for the two Maxwell scalars $\varphi
_{0}$ and $\varphi _{2}$.

For several problems, e.g. the calculation of energy and angular-momentum
outflux to infinity, knowledge of the Teukolsky variables (e.g. $\psi _{4}$
or $\varphi _{2}$) is sufficient. However, there are problems for which one
needs the full perturbation field (i.e. the MP $h_{\alpha \beta
}$ in the gravitational case, and $A_{\alpha }$ -- or alternatively the full
tensor field $F_{\alpha \beta }$ -- in the electromagnetic case). This
includes, for example, the calculation of gravitational or electromagnetic
self force acting on a point-like particle orbiting a spinning BH.

In principle, each of the Weyl scalars $\psi _{0}$ and $\psi _{4}$ contains
the full information on the gravitational perturbation in vacuum \cite{t2}
(up to a few non-radiative degrees of freedoms, e.g. infinitesimal
changes in the BH's mass and spin). Chrzanowski \cite{ch} developed a
procedure which allows the determination of the general homogeneous (i.e.
vacuum) solution for the MP $h_{\alpha \beta }$, by applying a certain
differential operator to the homogeneous solutions for $\psi _{0}$ or $\psi
_{4}$. It was later shown \cite{wald}, however, that this operator, when
applied to a particular solution $\psi _{0}$ or $\psi _{4}$ yields a
result $h_{\alpha \beta }$ which is a valid vacuum solution, but yet it
represents a {\em physically different} gravitational perturbation.
Let us rephrase this in a more explicit manner: 
Consider a vacuum gravitational perturbation
characterized by a particular function $\psi _{4}$. Then, there exists a
certain function $\Psi $, from which $h_{\alpha \beta }$ can be constructed
by applying Chrzanowski's differential operator. This function $\Psi $
satisfies the same Teukolsky equation as the function $\rho ^{-4}\psi _{4}$
(where $\rho $ is a certain quantity defined below), but yet $\Psi $ does 
{\em not} coincide with the quantity $\rho ^{-4}\psi _{4}$ of the
gravitational perturbation under consideration.

The same situation occurs in the case of electromagnetic perturbations.
Here, too, the full information about the (radiative part of the)
electromagnetic perturbation is encoded in each of the Maxwell scalars $
\varphi _{0}$ and $\varphi _{2}$ \cite{t2}. Chrzanowski's method \cite{ch}
allows the determination of the general, homogeneous solution for $A_{\alpha
}$ by applying a certain differential operator to the homogeneous solutions
for $\varphi _{0}$ or $\varphi _{2}$. However, this procedure, when applied
to a particular solution $\varphi _{0}$ or $\varphi _{2}$, yields a vacuum
solution $A_{\alpha }$ which represents a physically different
electromagnetic perturbation \cite{wald}.

In view of the above, the problem of constructing the MP $h_{\alpha \beta }$
(the four-potential $A_{\alpha }$) from $\psi _{0}$ or $\psi _{4}$ ($\varphi
_{0}$ or $\varphi _{2}$) includes two stages: First, construct the potential 
$\Psi $ from $\psi _{0}$ or $\psi _{4}$ ($\varphi _{0}$ or $\varphi _{2}$),
and second, construct $h_{\alpha \beta }$ ($A_{\alpha }$) from $\Psi $. The
second part is well known --- this is Chrzanowski's procedure \cite{ch}. The
goal of the present paper is to address the first part, namely, the
determination of $\Psi $ from $\psi _{0}$ (or from $\varphi _{0}$ in the
electromagnetic case).\footnote{
We shall restrict attention in this paper to the construction of $h_{\alpha
\beta }$ or $A_{\alpha }$ in the {\em ingoing} radiation gauge from $\psi
_{0}$ or $\varphi _{0}$, respectively. The analogous problem of constructing 
$h_{\alpha \beta }$ or $A_{\alpha }$ in the outgoing radiation gauge from $
\psi _{4}$ or $\varphi _{2}$ may be
treated in a similar way.} This problem was recently addressed, for
gravitational perturbations, by Lousto and Whiting (LW) \cite{LW} in the
case of a Schwarzschild BH. Here we provide the solution to this problem in
the Kerr case (in the frequency domain).

We shall consider here two different physical situations: (i) pure
gravitational or electromagnetic waves (i.e. a vacuum perturbation in the
entire spacetime), and (ii) gravitational (or electromagnetic) perturbations
produced by a (charged) object orbiting the BH. The first problem is fairly
simple, but the second one, that of perturbations with sources, is a bit
more involved. The explicit solution for $\Psi $ in this case of
inhomogeneous perturbations, which is the primary goal of this paper, is
summarized in sections \ref{sec8} (gravitational case) and \ref{sec9}
(electromagnetic case) below.

The equations relating the potential $\Psi $ to the relevant Teukolsky
variables were derived by Wald \cite{wald} for a general algebraically
special, vacuum, background spacetime. The reduction of these equations to
the Kerr case is given in Ref.\ \cite{CK} for the electromagnetic case and in
Ref.\ \cite{LW} for the gravitational case. Our goal is to determine $\Psi $
by solving these equations. For either the gravitational or electromagnetic
case, there are two such differential equations relating $\Psi $ to the
Teukolsky variables: a {\em radial equation} (i.e. one including $r$
-derivatives), which relates $\Psi $ to $\psi _{0}$ or $\varphi _{0}$, and
an {\em angular equation} (i.e. one including $\theta $-derivatives),
relating $\Psi $ to $\psi _{4}$ or $\varphi _{2}$. In Ref.\ \cite{LW} LW
elaborated on the angular equation [Eq.\ (2.7) therein], and constructed its
solution in the Schwarzschild case (for gravitational perturbations). Here
we shall elaborate on the radial equation [Eq.\ (2.6) therein, or its
electromagnetic counterpart]. This in fact turns out to be a simple ordinary
differential equation, which is not difficult to solve even in the Kerr case.

The MP $h_{\alpha \beta }$ and four-potential $A_{\alpha }$ constructed via
Chrzanowski's method are given in the ingoing (or outgoing) 
radiation gauge \cite{ch}. Barack and Ori \cite{BO}
recently investigated the local asymptotic behavior of the radiation-gauge 
$h_{\alpha \beta }$ (either the ingoing or outgoing one) near a point particle,
by locally integrating the equations defining this gauge. They found that
in this gauge $h_{\alpha \beta }$ cannot be well defied all around the
particle; Instead, there is a line of singularity that emerges from the
particle to either the ingoing or outgoing radial direction, over which 
$h_{\alpha \beta }$ diverges. (This line forms a $1+1$ dimensional
singularity set in spacetime.) One can choose to have a regular function 
$h_{\alpha \beta }$ at $r>r_{particle}$, where $r$ is the radial coordinate,
but this will inevitably lead to a line singularity at $r<r_{particle}$; and
vice versa. (Barack and Ori demonstrated this in the simplest case, i.e. a
static particle located at $r=r_{particle}$ in flat spacetime, but the
same situation should occur also for moving particles in Kerr.) Although the
analysis in Ref.\ \cite{BO} was restricted to the gravitational case, it is
easily extended to the electromagnetic case as well. It shows that the
radiation-gauge $A_{\alpha }$ also has a line singularity, either at 
$r>r_{particle}$ or at $r<r_{particle}$.

The solution constructed here provides an independent demonstration to this
pathology of the radiation-gauge quantities $h_{\alpha \beta }$ and 
$A_{\alpha }$ near a point source. Throughout this paper we shall assume that
the source is confined to a range $r_{\min }\leq r\leq r_{\max }$ (consider
e.g. a point mass moving on an elliptical or circular orbit). In both the
electromagnetic and gravitational cases (and for either the ingoing or
outgoing radiation gauge), one may choose to integrate the equations
governing $\Psi $ from $r>r_{\max }$ towards smaller $r$ values. Then $\Psi $
is perfectly regular at $r>r_{\max }$; but it turns out that at $r<r_{\max }$,
$\Psi $ is irregular on an outgoing null ray emerging form the particle
inwardly (i.e. in the past direction). Alternatively one may integrate these
equations from $r<r_{\min }$ towards larger $r$ values, in which case $\Psi $
is perfectly regular at $r<r_{\min }$ but at $r>r_{\min }$ it develops an
irregularity on an outgoing null ray emerging form the particle. This is
perfectly consistent with the above mentioned irregularity of the
radiation-gauge $h_{\alpha \beta }$ and $A_{\alpha }$, found earlier (for 
$h_{\alpha \beta }$) by Barack and Ori \cite{BO}. Throughout this paper we
shall mostly refer to the solution for $\Psi $ which is regular at 
$r>r_{\max }$ but has a line singularity at $r<r_{\max }$, which we shall
denote $\Psi ^{+}$ for concreteness. The analogous solution $\Psi ^{-}$
(which is regular at $r<r_{\min }$ but has a line singularity at $r>r_{\min
} $) may be constructed in a fully analogous manner, as briefly summarized
at the end of sections \ref{sec8} and \ref{sec9}. (In section \ref{sec10} we
briefly discuss the possible implications of this line singularity 
to the self-force problem.)

The two solutions $\Psi ^{+}$ and $\Psi ^{-}$ yield two different solutions
for the MP $h_{\alpha \beta }$ or the four-potential $A_{\alpha }$, both for
the ''same'' (e.g. the ingoing) radiation-gauge condition, which we denote 
$h_{\alpha \beta }^{+},A_{\alpha }^{+}$ and $h_{\alpha \beta }^{-},A_{\alpha
}^{-}$, correspondingly. To avoid confusion we emphasize that these two
solutions (for either $h_{\alpha \beta }$ or $A_{\alpha }$, and, say, in the
ingoing gauge) essentially 
represent {\em the same physical perturbation}, and they
differ by gauge. That is, the ingoing (or outgoing) radiation-gauge
condition does not completely fix the gauge. \footnote{It appears, though, 
that the ``$+$'' and ``$-$'' perturbations also differ by
some (non-gauge) non-radiative component, i.e. the so-called ``$l=0$'' and/or 
``$l=1$'' modes.}

Since there is full analogy between the gravitational and electromagnetic
cases, the detailed calculations presented in the next seven sections will
refer to the gravitational case only. The electromagnetic perturbations may
be treated exactly in the same manner. Only in section \ref{sec9} we shall
return to the electromagnetic case and present the final procedure of
constructing $\Psi $ for electromagnetic perturbations.

In section \ref{sec2} we give the basic field equations and establish some
notation. Section \ref{sec3} presents the basic calculation of $\Psi $ in
the case of pure gravitational waves, expressing it in terms of the modes of 
$\psi _{0}$. This result is then further simplified in section \ref{sec4},
by expressing both $\psi _{0}$ and $\Psi $ in terms of basis solutions (of
the relevant homogeneous Teukolsky equation) which admit a simple asymptotic
behavior at either the large-$r$ limit or at the event horizon (EH). In
section \ref{sec5} we develope the general homogeneous solution for Eq.\ (\ref
{e1}) below, the above mentioned ''radial equation'' relating $\Psi $ to 
$\psi _{0}$, which is a 4th-order differential equation. This general
homogeneous solution is required later for the construction of the relevant
inhomogeneous solution (for perturbations with sources).

In section \ref{sec6} we address the physical situation which provides the
main motivation for this paper, namely, gravitational perturbations produced
by sources (e.g. a point particle, or an extended object, in orbit around a
Kerr BH). We construct the solution for $\Psi $ (more specifically, $\Psi
^{+}$) in this case, using the general homogeneous solution constructed
earlier in section \ref{sec5}. In the first stage the potential $\Psi $ is
expressed in terms of the inhomogeneous mode functions of $\psi _{0}$. Then
we further simplify the solution, expressing $\Psi $ directly in terms of
the mode functions of the energy-momentum distribution in spacetime. (In
both stages we also use the homogeneous basis modes of $\psi _{4}$ in our
expression for $\Psi $). Whereas our detailed construction refers to $\Psi
^{+}$, the construction of $\Psi ^{-}$ proceeds in a fully analogous manner,
and we provide the final result for $\Psi ^{-}$ as well. Some details of the
calculations are given in Appendices A and B.

In section \ref{sec7} we study the domain of validity of the solution $\Psi
^{+}$ (and similarly of $\Psi ^{-}$). For a point particle we find that 
$\Psi ^{+}$ is regular everywhere except at a $(1+1)$ surface spanned by
outgoing principal null geodesics emanating from the particle's worldline in
the small-$r$ (i.e. past) direction. We denote this surface $\Sigma ^{+}$.
For an extended object, the solution $\Psi ^{+}$ (which is typically regular
throughout) is valid everywhere except in a domain denoted (again) $\Sigma
^{+}$, which is now a four-dimensional set (the definition of which is
provided therein). In a fully analogous manner, the other solution $\Psi
^{-} $ is, for an extended object, valid everywhere except in a
four-dimensional set $\Sigma ^{-}$; and at the point-like limit $\Sigma ^{-}$
degenerates to a $(1+1)$ surface spanned by outgoing principal null
geodesics emanating from the particle's worldline in the large-$r$ (i.e.
future) direction, with $\Psi ^{-}$ becoming irregular on $\Sigma ^{-}$.

Section \ref{sec8} provides a summary of the construction of $\Psi $, in the
gravitational case, for the benefit of the reader who wishes to implement
this method in practical calculations. Then, in section \ref{sec9} we return
to the electromagnetic problem and summarize the procedure of constructing 
$\Psi $ in this case, leaving many details of the derivation to Appendix C.
Finally in section \ref{sec10} we give some concluding remarks.

\section{Preliminaries}

\label{sec2}

Consider the spacetime of a Kerr BH with mass $M$ and specific
angular-momentum $a$. We shall use the standard Boyer-Lindquist coordinates 
($t,r,\theta ,\varphi $), and following Teukolsky \cite{t0} we denote $\Delta
\equiv r^{2}-2Mr+a^{2}$, $\Sigma \equiv r^{2}+a^{2}\cos ^{2}\theta $, and 
$\rho \equiv -1/(r-ia\cos \theta )$. The Newman-Penrose Weyl scalars $\psi
_{0}$ and $\psi _{4}$ (corresponding to $s=+2$ and $s=-2$, respectively)
satisfy two decoupled master equations. Defining 
\[
\psi _{+2}\equiv \psi _{0},\,\psi _{-2}\equiv \rho ^{-4}\psi _{4}\,, 
\]
we may formally write the two master equations as 
\begin{equation}
W_{\pm 2}\left[ \psi _{\pm 2}\right] =4\pi \Sigma T_{\pm 2}\,,  \label{m04}
\end{equation}
where $W_{\pm 2}$ is the second-order partial differential operator 
\begin{eqnarray}
&&W_{s}\equiv {-\Delta ^{-s}\partial _{r}[\Delta ^{s+1}\partial
_{r}]+[(r^{2}+a^{2})^{2}{/}\Delta -a^{2}\sin ^{2}\theta ]\,\partial _{tt}} 
\nonumber \\
&&{+\,4aMr\Delta }^{-1}{\partial _{\varphi \,t}+\,[a^{2}{/}\Delta -\sin
^{-2}\theta ]}\partial _{\varphi \varphi }  \nonumber \\
&&{-{\sin }^{-1}{\theta \,\partial _{\theta }[\sin \theta \,\partial
_{\theta }]}-2s\,[a(r-M){/}\Delta +i\,\cos \theta \sin ^{-2}\theta ]}
\partial _{\varphi }  \nonumber \\
&&{-2s\,[M(r^{2}-a^{2}){/}\Delta -r-ia\cos \theta ]\partial _{t}+(s^{2}\cot
^{2}\theta -s)\,,}  \label{meq}
\end{eqnarray}
and $T_{\pm 2}$ is the corresponding energy-momentum source term given
explicitly in Refs.\ \cite{t0}\cite{t1}. Teukolsky further showed that these
two Weyl scalars may be decomposed as
\begin{equation}
\psi _{\pm 2}=\sum_{\lambda m\omega }R_{\pm 2}^{\lambda m\omega }(r)S_{\pm
2}^{\lambda m\omega }(\theta )e^{i(m\varphi -\omega t)}\,,  \label{m5}
\end{equation}
where $R_{\pm 2}^{\lambda m\omega }$ and $S_{\pm 2}^{\lambda m\omega }$ are,
respectively, solutions of the radial and angular Teukolsky equations (given
below).\footnote{
In case the spectrum is continuous, the sum over $\omega $ should be
replaced by an integral.} The source terms are expanded in a similar manner: 
\begin{equation}
4\pi \Sigma T_{\pm 2}=\sum_{\lambda m\omega }T_{\pm 2}^{\lambda m\omega
}(r)S_{\pm 2}^{\lambda m\omega }(\theta )e^{i(m\varphi -\omega t)}\,.
\end{equation}
The radial and angular ordinary differential equations take the form 
\begin{equation}
P_{\pm 2}^{\lambda m\omega }\left[ R_{\pm 2}^{\lambda m\omega }(r)\right]
=T_{\pm 2}^{\lambda m\omega }(r)\,  \label{Tr}
\end{equation}
and 
\begin{equation}
\Theta _{\pm 2}^{\lambda m\omega }\left[ S_{\pm 2}^{\lambda m\omega }(\theta
)\right] =0\,.  \label{Ts}
\end{equation}
Here $P_{\pm 2}^{\lambda m\omega }$ and $\Theta _{\pm 2}^{\lambda m\omega }$
are second-order linear differential operators, given by 
\begin{equation}
P_{s}^{\lambda m\omega }\equiv {\Delta ^{-s}\partial _{r}[\Delta
^{s+1}\partial _{r}]+V}_{r}{(r)\,}  \label{Trexp}
\end{equation}
and 
\[
\Theta _{s}^{\lambda m\omega }\equiv {\,{\sin }^{-1}{\theta \,\partial
_{\theta }[\sin \theta \,\partial _{\theta }]}+V}_{\theta }{(\theta )\,,} 
\]
where the potentials are 
\begin{eqnarray*}
{V}_{r}{(r)} &=&{[(r^{2}+a^{2})^{2}{\omega }^{2}}-{4aM\omega mr+a^{2}m^{2}}
\\
&&{+2iams(r-M)-2iM\omega s(r^{2}-a^{2})]\Delta }^{-1} \\
&&+2i{\omega sr-a^{2}{\omega }^{2}-\tilde{A}\,}
\end{eqnarray*}
and 
\begin{eqnarray*}
{V}_{\theta }{(\theta )} &=&{a^{2}{\omega }^{2}\cos }^{2}\theta -m^{2}/\sin
^{2}\theta -2a\omega s{\cos }\theta \\
&&-2ms{\cos }\theta /\sin ^{2}\theta -{s^{2}\cot ^{2}\theta +s+\tilde{A}\,\,.
}
\end{eqnarray*}
The parameter ${\tilde{A}}$ is Teukolsky's \cite{t0} separation constant,
which we write as 
\[
{\tilde{A}=\lambda -}s-|s|\,, 
\]
where $\lambda $ is the separation constant used by Chandrasekhar \cite
{chandra} (often denoted ${\not{\lambda}}$ there). The parameter $\lambda $
runs over all eigenvalues of the angular Teukolsky equation (\ref{Ts}).
Throughout this paper we prefer to use the separation constant $\lambda $
rather than ${\tilde{A}}$, because the angular equations for $s=2$ and $s=-2$
have the same set of eigenvalues $\lambda $ \cite{chandra} 
(which is not the case for ${\tilde{A}}$). \footnote{
In the special case $a\omega =0$, the separation constant $\lambda $ becomes 
$l(l+1)-s^{2}+|s|$; hence, common $\lambda $ also means common $l$.} 
Also, this will allow an easier connection with Chandrasekhar's formalism.

Our goal is to construct the MP. In a vacuum spacetime (i.e. $T_{\pm 2}=0$),
the MP in the ingoing radiation gauge can be derived from a potential $\Psi
_{IRG}$ \cite{ch}, by applying to the latter a certain second-order
differential operator \cite{operator}. We shall formally denote this
differential operator by $\Pi _{IRG}$, namely, 
\begin{equation}
h_{IRG}=\Pi _{IRG}\left[ \Psi _{IRG}\right] \,,  \label{h}
\end{equation}
where $h_{IRG}$ denotes the MP in the ingoing radiation gauge (for brevity
we shall often omit the spacetime indices of the MP). Similarly, the MP in the
outgoing radiation gauge can be obtained from another potential $\Psi _{OUT}$,
through another differential operator $\Pi _{OUT}$ \cite{operator}, namely 
\[
h_{ORG}=\Pi _{ORG}\left[ \Psi _{ORG}\right] \,. 
\]
In this paper we shall only consider the case of ingoing radiation gauge,
but the potential $\Psi _{ORG}$ (and hence the MP in the outgoing radiation
gauge) may be constructed in a fully analogous manner. For brevity we shall
use here the notation $\Psi \equiv \Psi _{IRG}$ (this variable is denoted $
\psi _{G}$ in Ref.\ \cite{wald}), $\Pi \equiv \Pi _{IRG}$, and $h_{\alpha
\beta }\equiv h_{IRG}$, hence 
\begin{equation}
h_{\alpha \beta }=\Pi \left[ \Psi \right] \,.  \label{MP}
\end{equation}

The function $\Psi $ has to be a solution of the vacuum Teukolsky equation
for $\psi _{-2}$ \cite{wald}, namely, 
\begin{equation}
W_{-2}\left[ \Psi \right] =0\,.  \label{eqp}
\end{equation}
In addition, it must satisfy the following differential equation \cite{wald}
\cite{LW}

\begin{equation}
\psi _{0}=D^{4}\left[ \bar{\Psi}\right] \,,  \label{e1}
\end{equation}
where throughout this paper an over-bar denotes complex conjugation. Here $D$
is the differential operator 
\[
D=l^{\mu }\partial _{\mu }=\frac{r^{2}+a^{2}}{\Delta }\partial _{t}+\partial
_{r}+(a/\Delta )\partial _{\varphi }\,, 
\]
where $l^{\mu }$ is the standard outgoing Kinnersley's tetrad vector (see
e.g. Ref.\ \cite{t0}). We use here the abbreviated notation $D^{4}\equiv DDDD$
, and the same for other operators used below.

Our goal in this paper is to construct the function $\Psi $ that satisfies
equations (\ref{eqp}) and (\ref{e1}). This will allow the construction of 
$h_{\alpha \beta }$, the MP in the ingoing radiation gauge, through Eq.\ (\ref
{MP}). We shall first consider the case of pure vacuum gravitational waves
in the entire spacetime. In this case we assume that $\psi _{0}$ is known
(it encodes the information on the gravitational waves). Subsequently we
shall consider the case of gravitational perturbations produced by a point
particle (or any other matter source) moving in the Kerr spacetime. In this
case we shall assume that $T_{+2}$, the $s=+2$ energy-momentum source term,
is known.

\section{Pure gravitational waves}

\label{sec3}

We now consider the case of pure vacuum gravitational waves, namely, $T_{\pm
2}=0$ in the entire spacetime. The information about the gravitational waves
is given by means of the Weyl scalar $\psi _{0}$. Since we are dealing here
with linear perturbations, it will be sufficient to consider a particular
mode $\lambda m\omega $ of $\psi _{0}$; The entire perturbation is then
obtained by a superposition. Thus, we now assume that $\psi _{0}$ takes the
decomposed form

\begin{equation}
\psi _{0}=R_{+2}^{\lambda m\omega }(r)S_{+2}^{\lambda m\omega }(\theta
)e^{i(m\varphi -\omega t)}\,,  \label{3}
\end{equation}
and the radial function satisfies the vacuum radial equation: 
\begin{equation}
P_{+2}^{\lambda m\omega }\left[ R_{+2}^{\lambda m\omega }(r)\right] =0\,.
\label{vacuum}
\end{equation}
We shall now use the Teukolsky-Starobinsky relations to show that the
desired solution of Eqs.\ (\ref{eqp},\ref{e1}) is 
\begin{equation}
\bar{\Psi}=p^{-1}\Delta ^{2}(D^{\dagger })^{4}[\Delta ^{2}\psi _{0}]\,,
\label{4}
\end{equation}
where $p$ is a constant to be determined later, and $D^{\dagger }$ is the
differential operator 
\[
D^{\dagger }=-\frac{r^{2}+a^{2}}{\Delta }\partial _{t}+\partial
_{r}-(a/\Delta )\partial _{\varphi }\,. 
\]
For a single $\lambda m\omega $ mode we define the ''reduced'' operators 
\[
D_{m\omega }=\partial _{r}+iK/\Delta \,\,,\,D_{m\omega }^{\dagger }=\partial
_{r}-iK/\Delta \,, 
\]
where 
\[
K\equiv am-(r^{2}+a^{2})\omega \,, 
\]
such that for any functions $f(r)$ and $g(\theta )$, 
\[
D\left[ f(r)g(\theta )e^{i(m\varphi -\omega t)}\right] =g(\theta
)e^{i(m\varphi -\omega t)}D_{m\omega }\left[ f(r)\right] \,, 
\]
and the same relation holds between the operators $D^{\dagger },D_{m\omega
}^{\dagger }$. (Note that $D_{m\omega }$ and $D_{m\omega }^{\dagger }$ are
the same as the operators ''${\cal D}_{0}$'' and ''${\cal D}_{0}^{\dagger }$
'', respectively, in Chandrasekhar's notation \cite{chandra}.) Using the
decomposition

\begin{equation}
\bar{\Psi}=\hat{R}_{-2}^{\lambda m\omega }S_{+2}^{\lambda m\omega }(\theta
)e^{i(m\varphi -\omega t)}\,,  \label{14}
\end{equation}
Eq.\ (\ref{4}) becomes 
\begin{equation}
\hat{R}_{-2}^{\lambda m\omega }=p^{-1}\Delta ^{2}(D_{m\omega }^{\dagger
})^{4}\left[ \Delta ^{2}R_{+2}^{\lambda m\omega }(r)\right] \,,  \label{15}
\end{equation}
and Eq.\ (\ref{e1}) now reduces to 
\begin{equation}
R_{+2}^{\lambda m\omega }(r)=(D_{m\omega })^{4}[\hat{R}_{-2}^{\lambda
m\omega }]\,.  \label{11}
\end{equation}

Let us first verify that $\Psi $ [the complex conjugate of
Eq.\ (\ref{4})] satisfies Eq.\ (\ref{eqp}). 
The Teukolsky-Starobinsky relations (see e.g. \cite
{chandra}) imply that the radial function $\hat{R}_{-2}^{\lambda m\omega
}(r) $ constructed in Eq.\ (\ref{15}) is a solution of the $s=-2$ radial
vacuum equation, namely 
\begin{equation}
P_{-2}^{\lambda m\omega }[\hat{R}_{-2}^{\lambda m\omega }(r)]=0\,.
\label{16}
\end{equation}
The complex conjugate of the radial function $\hat{R}_{-2}^{\lambda m\omega
} $ is a radial vacuum solution with the same $\lambda $ and $s=-2$, but
with negative sign for $m$ and $\omega $, which we denote $\hat{R}
_{-2}^{\lambda ,-m,-\omega }$. The angular Teukolsky equation is real, and
(holding $\lambda $ fixed) is invariant under the simultaneous change of
signs of $s,m,\omega $. Hence, $S_{+2}^{\lambda m\omega }(\theta )$ is a
real function which is also a solution to the angular Teukolsky equation 
(\ref{Ts}) with $-m,-\omega $, and $s=-2$; and correspondingly we may write
it as $cS_{-2}^{\lambda ,-m,-\omega }$, where $c$ is a constant (whose value
is unimportant to us). We find that 
\begin{equation}
\Psi =c\hat{R}_{-2}^{\lambda ,-m,-\omega }(r)S_{-2}^{\lambda ,-m,-\omega
}(\theta )e^{-i(m\varphi -\omega t)}\,.
\end{equation}
Thus, $\Psi $ is indeed a solution to the $s=-2$ vacuum Teukolsky equation (\ref
{eqp}) -- a solution characterized by the set of indices $\lambda
,-m,-\omega $.

We still need to check that $\hat{R}_{-2}^{\lambda m\omega }$ constructed in
Eq.\ (\ref{15}) satisfies Eq.\ (\ref{11}) [this would in turn imply that 
the expression 
(\ref{4}) satisfies Eq.\ (\ref{e1})], and to determine the constant $p$. In
fact all we need to show is that
\[
(D_{m\omega })^{4}\left[ \Delta ^{2}(D_{m\omega }^{\dagger })^{4}[\Delta
^{2}R_{+2}^{\lambda m\omega }(r)]\right] 
\]
is a constant multiple of $R_{+2}^{\lambda m\omega }$. This follows
immediately by applying the two parts of Theorem 1 in Chapter 9 of Ref.\ \cite
{chandra}. Consequently, there exists a constant $p$ such that 
\begin{equation}
R_{+2}^{\lambda m\omega }(r)=p^{-1}(D_{m\omega })^{4}\left[ \Delta
^{2}(D_{m\omega }^{\dagger })^{4}[\Delta ^{2}R_{+2}^{\lambda m\omega
}(r)]\right] \,.  \label{RR}
\end{equation}
From the analysis therein it becomes obvious \footnote{
This may easily be deduced by applying the operator $\Delta ^{2}(D_{m\omega
}^{\dagger })^{4}\Delta ^{2}$ to both sides of Eq.\ (\ref{RR} ), and then
using Chandrasekhar's theorem 1, as well as his Eq.\ (43) (both in Chapter 9
of Ref.\ \cite{chandra}).} that $p$ is the real constant

\begin{eqnarray}
p &=&\lambda ^{2}(\lambda +2)^{2}-8\omega ^{2}\lambda [\alpha ^{2}(5\lambda
+6)-12a^{2}]  \nonumber \\
&&+144\omega ^{4}\alpha ^{4}+144\omega ^{2}M^{2}\,,  \label{p}
\end{eqnarray}
where $\alpha ^{2}\equiv a^{2}-am/\omega $. Note that the coefficient $p$
depends on the mode. \footnote{
It is assumed that $p$ is finite and non-vanishing for all real $\omega $.}

As was mentioned in section \ref{sec1}, the potential $\Psi$ must also 
satisfy an angular differential equation, i.e. Eq.\ (2.7) in Ref.\ \cite{LW}.
The compliance of the above constructed vacuum solution with this additional 
equation is guaranteed by virtue of the following considerations: 
(i) According to the analysis in Ref.\ \cite{wald}, there must exist a 
solution to the three simultaneous equations [i.e. Eqs.\ 
(\ref{eqp}) and (\ref{e1}), and the angular equation]; and 
(ii) The solution (\ref{14},\ref{15}) is the {\it unique} 
solution to the simultaneous equations (\ref{eqp}) and (\ref{e1}). 
For, any nontrivial solution to the homogeneous part of Eq.\ (\ref{e1}) 
will violate Eq.\ (\ref{eqp}) (one can easily verify this,
based on the general homogeneous solution to  Eq.\ (\ref{e1}), constructed 
in section \ref{sec5} below).

\section{Further simplification of the vacuum solution}

\label{sec4}

Equations (\ref{14},\ref{15}) provide the full solution for $\bar{\Psi}$. It
is possible, however, to construct a simpler and more explicit expression
for the radial function $\hat{R}_{-2}^{\lambda m\omega }$. The latter
satisfies Eq.\ (\ref{16}), which is the vacuum Teukolsky equation for the $
s=-2$ radial function $R_{-2}^{\lambda m\omega }$. Since this is a
second-order ordinary differential equation, its general solution may be
spanned by any pair of independent solutions. Let $R_{\pm 2}^{\lambda
m\omega (a)}$ and $R_{\pm 2}^{\lambda m\omega (b)}$ be such two pairs of
independent solutions (one pair for $s=+2$ and one for $s=-2$). Let $H$ denote
the operator which maps a vacuum $s=+2$ radial function $R_{+2}^{\lambda
m\omega }$ into the corresponding function $\hat{R}_{-2}^{\lambda m\omega }$
of Eq.\ (\ref{15}), namely, 
\[
H\equiv p^{-1}\Delta ^{2}(D_{m\omega }^{\dagger })^{4}\Delta ^{2}\,. 
\]
For each mode $\lambda m\omega $ there must exist a constant $2\times 2$
matrix $C_{ij}$ such that 
\begin{equation}
H[R_{+2}^{\lambda m\omega (i)}(r)]=C_{ij}R_{-2}^{\lambda m\omega (j)}(r)\,,
\label{basis}
\end{equation}
where $i,j$ run over the two basis states $a$ and $b$. The problem thus
reduces to the determination of the four constants $C_{ij}$.

There are two preferred bases, however, for which this matrix becomes
diagonal and especially easy to calculate. One such basis is the pair of
solutions characterized by positive and negative exponents of $r_{*}$ at
large $r$. Here $r_{*}$ is a function of $r$, defined by 
\begin{equation}
dr_{*}/dr=(r^{2}+a^{2})/\Delta \,  \label{r*}
\end{equation}
(and given explicitly below). Note that $r_{*}\to \infty $ as $r\to \infty $.
 The other basis is that of positive and negative exponents of $r_{*}$ at
the EH (where $r_{*}\to -\infty $). These two bases are also preferable for
the physical interpretation of the solution, and for its construction via
a Green function (described in section \ref{sec6} below). The asymptotic 
behavior of the
vacuum radial Teukolsky functions are given in e.g.
Ref.\ \cite{t2} for all values of $s$, 
both at the limit of large $r$ and at the EH.

In what follows we shall describe the application of Eq.\ (\ref{basis}), and
the determination of the required coefficients, for these two special bases.

\subsection{Large-$r$ asymptotic behavior}

Considering the large-$r$ asymptotic behavior of the vacuum radial functions 
$R_{+2}^{\lambda m\omega }(r)$ and $R_{-2}^{\lambda m\omega }(r)$, we may
take the two basic solutions (for each $s$) to be those of positive and
negative exponents of $r_{*}$. These two solutions take the asymptotic form 
\begin{equation}
R_{+2}^{\lambda m\omega (in)}\cong e^{-i\omega
r_{*}}/r\,\,,\,\,\,\,\,\,R_{+2}^{\lambda m\omega (out)}\cong e^{i\omega
r_{*}}/r^{5}\,  \label{in}
\end{equation}
and 
\begin{equation}
R_{-2}^{\lambda m\omega (in)}\cong e^{-i\omega
r_{*}}/r\,\,,\,\,\,\,\,\,R_{-2}^{\lambda m\omega (out)}\cong e^{i\omega
r_{*}}r^{3}\,  \label{out}
\end{equation}
(see \cite{t2}, and recall the factor $\rho ^{-4}\propto r^{4}$ in 
the above
definition of $\psi _{-2}$). To avoid confusion we emphasize that the
basis solutions $R_{\pm 2}^{\lambda m\omega (in,out)}$ are defined to be the 
{\em exact} solutions of the corresponding radial equations, which satisfy
the asymptotic form (\ref{in},\ref{out}) at the leading order in $1/r$ (the
same remark applies to the event-horizon basis functions defined below.)

One can easily verify that the operators $D_{m\omega },D_{m\omega }^{\dagger
}$ do not mix positive and negative exponents or $r_{*}$. Therefore, 
$H[R_{+2}^{\lambda m\omega (in)}]$ and $H[R_{+2}^{\lambda m\omega (out)}]$
must take the simple forms 
\begin{equation}
H[R_{+2}^{\lambda m\omega (in)}]=C^{(in)}R_{-2}^{\lambda m\omega (in)}
\label{cin}
\end{equation}
and

\begin{equation}
H[R_{+2}^{\lambda m\omega (out)}]=C^{(out)}R_{-2}^{\lambda m\omega (out)}\,,
\label{cout}
\end{equation}
and the problem reduces to the determination of the two constants $C^{(in)}$
and $C^{(out)}$. These constants may be determined from the large-$r$
asymptotic form of Eqs.\ (\ref{15}) or (\ref{11}). Ignoring terms of higher
order in $1/r$, we have 
\[
D_{m\omega }\cong \partial _{r}-i\omega \,,\,D_{m\omega }^{\dagger }\cong
\partial _{r}+i\omega \,\,, 
\]
and $\Delta \cong r^{2}$.

In principle, both Eqs.\ (\ref{15}) and (\ref{11}) may be used for the
determination of each of the coefficients $C^{(in)},C^{(out)}$. Notice,
however, that when $D_{m\omega }$ acts on $R_{\pm 2}^{\lambda m\omega (out)}$
and $D_{m\omega }^{\dagger }$ on $\Delta ^{2}R_{\pm 2}^{\lambda m\omega
(in)} $, the leading-order term proportional to $\omega $ cancels out.
Therefore, in these cases the operator effectively decreases the powers of 
$r $ (by $1$ at least), as $\partial _{r}$ differentiates this power of $r$.
This leads to a complication, because then we cannot ignore the higher-order
terms (in $1/r$) in the operators $D_{m\omega },D_{m\omega }^{\dagger }$
and in $\Delta $, and also the higher-order terms in the basis solutions $
R_{\pm 2}^{\lambda m\omega (in,out)}$. On the other hand, no such
cancelation of the leading order term occurs when $D_{m\omega }$ acts on 
$R_{\pm 2}^{\lambda m\omega (in)}$ and $D_{m\omega }^{\dagger }$ on $R_{\pm
2}^{\lambda m\omega (out)}$. Instead, we get 
\[
D_{m\omega }[R_{\pm 2}^{\lambda m\omega (in)}]\cong -2i\omega R_{\pm
2}^{\lambda m\omega (in)}\, 
\]
and 
\[
\,D_{m\omega }^{\dagger }[R_{\pm 2}^{\lambda m\omega (out)}]\cong 2i\omega
R_{\pm 2}^{\lambda m\omega (out)}\,. 
\]
It will therefore be convenient to calculate $C^{(in)}$ from Eq.\ (\ref{11})
and $C^{(out)}$ from Eq.\ (\ref{15}), by substituting in these equations 
\begin{equation}
R_{+2}^{\lambda m\omega }=R_{+2}^{\lambda m\omega (a)}\,,\,\,
\hat{R}_{-2}^{\lambda m\omega }
=C^{(a)}R_{-2}^{\lambda m\omega (a)}  \label{subst}
\end{equation}
(with ''$a$'' standing for either ''in'' or ''out'', as appropriate).
Equation (\ref{11}) then becomes 
\[
R_{+2}^{\lambda m\omega (in)}\cong 16\omega ^{4}C^{(in)}R_{-2}^{\lambda
m\omega (in)}\,, 
\]
and Eq.\ (\ref{15}) reads 
\[
C^{(out)}R_{-2}^{\lambda m\omega (out)}\cong 16\omega
^{4}p^{-1}r^{8}R_{+2}^{\lambda m\omega (out)}\,. 
\]
Since Eqs.\ (\ref{in},\ref{out}) imply 
\[
R_{-2}^{\lambda m\omega (in)}\cong R_{+2}^{\lambda m\omega
(in)}\,\,,\,\,R_{-2}^{\lambda m\omega (out)}\cong r^{8}R_{+2}^{\lambda
m\omega (out)}\,, 
\]
we obtain 
\begin{equation}
C^{(in)}=1/(16\omega ^{4})\,,\,C^{(out)}=16\omega ^{4}/p\,.  \label{cf}
\end{equation}

\subsection{Event-Horizon asymptotic behavior}

In a completely analogous manner, we can use for the expansion of 
$\,R_{+2}^{\lambda m\omega }$ and $\hat{R}_{-2}^{\lambda m\omega }$ basis
solutions characterized by either negative or positive exponents of $r_{*}$
at the EH (the latter corresponds to $r_{*}\to -\infty $). These basis
solutions take the asymptotic form 
\begin{equation}
R_{+2}^{\lambda m\omega (down)}\cong \Delta
^{-2}e^{-ikr*}\,\,,\,\,\,\,\,\,R_{+2}^{\lambda m\omega (up)}\cong
e^{ikr*}\,  \label{hor+}
\end{equation}
and 
\begin{equation}
R_{-2}^{\lambda m\omega (down)}\cong \Delta
^{2}e^{-ikr*}\,\,,\,\,\,\,\,\,R_{-2}^{\lambda m\omega (up)}\cong
e^{ikr*}\,  \label{hor-}
\end{equation}
(see \cite{t2}). Here $k\equiv \omega -ma/(2Mr_{+})$, where $r_{+}$ is the 
$r $ value at the event horizon, given by 
\[
r_{\pm }=M\pm (M^{2}-a^{2})^{1/2}\,. 
\]
In this case, again, one can verify that the operators $D_{m\omega
},D_{m\omega }^{\dagger }$ do not mix positive and negative exponents of 
$r_{*}$. Therefore, 
\begin{equation}
H[R_{+2}^{\lambda m\omega (down)}]=C^{(down)}R_{-2}^{\lambda m\omega (down)}
\label{hdown}
\end{equation}
and

\begin{equation}
H[R_{+2}^{\lambda m\omega (up)}]=C^{(up)}R_{-2}^{\lambda m\omega (up)}\,,
\end{equation}
and the problem reduces to the determination of the two constants 
$C^{(down)} $ and $C^{(up)}$.

The leading-order form of $\Delta =(r-r_{+})(r-r_{-})$ is 
\[
\Delta \cong q\delta r\,, 
\]
where $\delta r\equiv r-r_{+}$ and 
\[
q=r_{+}-r_{-}=2(M^{2}-a^{2})^{1/2}\,. 
\]
Correspondingly, the leading-order forms of $D_{m\omega }$ and $\,D_{m\omega
}^{\dagger }$ near the EH are 
\[
D_{m\omega }\cong \frac{2Mr_{+}}{\Delta }\left( \partial _{r_{*}}-ik\right)
\,,\,D_{m\omega }^{\dagger }\cong \frac{2Mr_{+}}{\Delta }\left( \partial
_{r_{*}}+ik\right) \,, 
\]
where we have used $r_{+}^{2}+a^{2}=2Mr_{+}$. However, when applying the
operators $D_{m\omega },D_{m\omega }^{\dagger }$ to the above basis
solutions, it is most convenient to view $r$ and $r_{*}$ in Eqs.\ (\ref{hor+},
\ref{hor-}) as two independent variables. In this context we have 
\[
D_{m\omega }\cong \partial _{r}+\frac{2Mr_{+}}{\Delta }\left( \partial
_{r_{*}}-ik\right) \,, 
\]
and 
\[
D_{m\omega }^{\dagger }\cong \partial _{r}+\frac{2Mr_{+}}{\Delta }\left
(\partial _{r_{*}}+ik\right) \,. 
\]
The above basis functions all take the general form $F(r)e^{\pm ikr*}$. For
such functions we have 
\[
D_{m\omega }\left[ Fe^{ikr*}\right] =F^{\prime }e^{ikr*}\,, 
\]
\[
D_{m\omega }\left[ Fe^{-ikr*}\right] =\left[ F^{\prime }-\frac{iw}{\Delta }
\right] e^{-ikr*}\,, 
\]
\[
D_{m\omega }^{\dagger }\left[ Fe^{ikr*}\right] =\left[ F^{\prime }+\frac{iw}
{\Delta }\right] e^{ikr*}\,, 
\]
\[
D_{m\omega }^{\dagger }\left[ Fe^{-ikr*}\right] =F^{\prime }e^{-ikr*}\,, 
\]
where $w=4kMr_{+}$ and a prime denotes $d/dr$. Note that when $D_{m\omega }$
acts on $R_{\pm 2}^{\lambda m\omega (up)}$ and $D_{m\omega }^{\dagger }$ on 
$\Delta ^{2}R_{+2}^{\lambda m\omega (down)}$, the leading-order term
proportional to $k$ cancels out, and we are left with higher-order terms
that take the lead, which is an inconvenient situation. For this reason we
shall calculate $C^{(down)}$ from Eq.\ (\ref{11}) and $C^{(up)}$ from Eq.\ 
(\ref{15}). Using again the substitution (\ref{subst}) (this time with ''$a$
'' standing for either ''down'' or ''up''), Eqs.\ (\ref{11}) and (\ref{15})
become, respectively, 
\begin{equation}
R_{+2}^{\lambda m\omega (down)}=C^{(down)}(D_{m\omega })^{4}[R_{-2}^{\lambda
m\omega (down)}]\,
\end{equation}
and 
\[
C^{(up)}R_{-2}^{\lambda m\omega (up)}=p^{-1}\Delta ^{2}(D_{m\omega
}^{\dagger })^{4}[\Delta ^{2}R_{+2}^{\lambda m\omega (up)}]\,. 
\]
For the first equation we need to calculate the quantity 
\begin{equation}
(D_{m\omega })^{4}[\Delta ^{2}e^{-ikr*}]\,.  \label{17}
\end{equation}
A straightforward calculation yields (at the leading order) 
\[
(D_{m\omega })^{4}[\Delta ^{2}e^{-ikr*}]\cong Q\Delta ^{-2}e^{-ikr*}\,, 
\]
where 
\[
Q=(w+2iq)(w+iq)w(w-iq)\,. 
\]
We thus find 
\begin{equation}
C^{(down)}=1/Q\,.  \label{cdown}
\end{equation}
For the second equation we need to calculate the quantity 
\[
(D_{m\omega }^{\dagger })^{4}[\Delta ^{2}e^{ikr*}]\,. 
\]
This is just the complex conjugate of expression (\ref{17}), and we find 
\[
(D_{m\omega }^{\dagger })^{4}[\Delta ^{2}e^{ikr*}]\cong \bar{Q}\Delta
^{-2}e^{ikr*}\,, 
\]
which leads to 
\begin{equation}
C^{(up)}=\bar{Q}/p\,.  \label{cup}
\end{equation}

\section{The general homogeneous solution to the 4th-order equation}

\label{sec5}

The equation (\ref{e1}) that determines the potential $\bar{\Psi}$ is an
inhomogeneous fourth-order linear differential equation.
In the last section we constructed the relevant inhomogeneous
solution of this equation in the case of pure vacuum perturbations, for each
mode of the source term $\psi _{0}$ 
[by ''relevant'' we refer here to the solution
that also solves the vacuum Teukolsky equation (\ref{eqp})]. Later we shall
also need the general solution of this fourth-order equation, in order to
construct the inhomogeneous solutions relevant to non-vacuum perturbations.
To this end, we shall now construct the {\em general homogeneous solution}
to Eq.\ (\ref{e1}).

This equation is in fact a trivial ordinary differential equation. Let us
denote by $\xi $ the null geodesics whose tangent is the null tetrad vector 
$l^{\mu }$ (namely, $\xi $ are the members of the outgoing principal null
congruence). Let $\gamma $ be an affine parameter along the geodesics $\xi $,
namely 
\[
l^{\mu }=\frac{dx^{\mu }}{d\gamma }(\xi )\,. 
\]
Then for any function $f(t,r,\theta ,\varphi )$, 
\[
D[f]\equiv l^{\mu }\partial _{\mu }f=\frac{df}{d\gamma }(\xi )\,. 
\]
The differential equation (\ref{e1}) thus reads 
\begin{equation}
\frac{d^{4}\bar{\Psi}}{d\gamma ^{4}}(\xi )=\psi _{0}\,\,.  \label{ord}
\end{equation}
Its general homogeneous solution is a third-order polynomial in $\gamma $,
whose four coefficients may be taken to be arbitrary functions of $\xi $: 
\begin{equation}
\bar{\Psi}=\sum_{i=0}^{3}b_{i}(\xi )\gamma ^{i}\,\,\,\,\,\,\,
\text{(homogeneous)}\,.  \label{generalt}
\end{equation}

We wish, however, to re-write this homogeneous solution more explicitly as a
function of the four spacetime coordinates. To this end we need to
explicitly parametrize the null geodesics $\xi $. From the definition of 
$l^{\mu }$, along each null geodesic $\xi $ we have 
\[
\frac{dr}{d\gamma }=1\,,\,\,\frac{d\theta }{d\gamma }=0\,,\,\,\frac{dt}
{d\gamma }=\frac{r^{2}+a^{2}}{\Delta }\,,\,\,\frac{d\varphi }{d\gamma }
=a/\Delta \,\,.\,\, 
\]
We choose the origin of $\gamma $ such that $\gamma =r$ along the geodesic.
Then $t,\theta ,\varphi $ along the geodesic are given by 
\begin{equation}
\theta =\theta _{0}\,,\,\,t=t_{0}+r_{*}(r)\,,\,\,\varphi =\varphi
_{0}+u(r)\,,  \label{parameters}
\end{equation}
where $\theta _{0},t_{0},\varphi _{0}$ are arbitrary constants, and $
r_{*}(r) $ and $u(r)$ are given by the two integrals 
\[
r_{*}(r)=\int \frac{r^{2}+a^{2}}{\Delta }dr\,,\,\,u(r)=\int \frac{a}{\Delta }
dr\,. 
\]
Specifically we take 
\begin{equation}
r_{*}(r)=r+\frac{r_{+}^{2}+a^{2}}{q}\ln (r-r_{+})-\frac{r_{-}^{2}+a^{2}}{q}
\ln (r-r_{-})  \label{r*d}
\end{equation}
and 
\begin{equation}
u(r)=(a/q)\ln [(r-r_{+})/(r-r_{-})]\,.  \label{u}
\end{equation}
The null geodesics $\xi $ are thus parametrized by the three parameters 
$\theta _{0}\equiv \theta $, $t_{0}\equiv t-r_{*}(r)$, and $\varphi
_{0}\equiv \varphi -u(r)$, and the above general solution takes the explicit
form 
\begin{equation}
\bar{\Psi}=\sum_{i=0}^{3}b_{i}(\theta _{0},t_{0},\varphi
_{0})r^{i}\,\,\,\,\,\,\,\,\,\,\,\,\,\,\,\,\,\,\text{(homogeneous)}\,,
\label{generalr}
\end{equation}
where $b_{i}$ are arbitrary functions of their arguments.

Later we shall also need the form of this general homogeneous solution in
the frequency domain. In order to comply with the decomposed form 
(\ref{14}), for a particular mode $\lambda m\omega $ the arbitrary 
functions $b_{i}$ must take the form 
\[
b_{i}=B_{i}S_{+2}^{\lambda m\omega }(\theta )e^{im(\varphi -u)}e^{-i\omega
(t-r_{*})}\,, 
\]
where $B_{i}$ are four arbitrary constants (for each mode). Correspondingly
the (homogeneous-solution) radial function $\hat{R}_{-2}^{\lambda m\omega }$
is given by 
\begin{equation}
\hat{R}_{-2}^{\lambda m\omega }(r)=e^{-i(mu-\omega
r_{*})}\sum_{i=0}^{3}B_{i}r^{i}\,\,\,\,\,\,\,\,\,\,\,\,\,\,\,\,\text{
(homogeneous)}\,.  \label{general1}
\end{equation}
One can easily verify that this solution indeed satisfies 
the homogeneous part of Eq.\ (\ref{11}), namely
\[
(D_{m\omega })^{4}[\hat{R}_{-2}^{\lambda m\omega
}]=0\,\,\,\,\,\,\,\,\,\,\,\,\,\,\,\text{(homogeneous)}\,. 
\]
To this end, it is sufficient to note that for any function $f(r)$, 
\begin{equation}
D_{m\omega }\left[ f(r)e^{-i(mu-\omega r_{*})}\right] =\frac{df}{dr}
e^{-i(mu-\omega r_{*})}\,.  \label{Diff}
\end{equation}

\section{Gravitational perturbations produced by sources}

\label{sec6}

Consider now gravitational waves produced by a point-like particle that
moves freely in a Kerr spacetime. For concreteness let us assume that the
orbit is confined to the range $r_{\min }\leq r\leq r_{\max }$ (but this
assumption may be relaxed, at least partially, as we discuss in section \ref
{sec10}). The orbit needs not be equatorial. Of special importance is the
case of a circular orbit, $r_{\min }=r_{\max }\equiv r_{0}$. Alternatively,
we may assume that the gravitational waves are produced by a finite-size mater
distribution. In this case, too, we shall assume that the matter is confined
to the range $r_{\min }\leq r\leq r_{\max }$.

In the formalism used here the single function $\Psi $ is required to
satisfy two differential equations -- Eq.\ (\ref{eqp}) and the inhomogeneous
equation (\ref{e1}). These equations are mutually-consistent in the case the
source term for Eq.\ (\ref{e1}) is a {\em vacuum} gravitational field $\psi
_{0}$, but otherwise we should expect to have an over-determination.
Therefore, in a spacetime with a matter source (either a finite-size or a
point-like source), we cannot expect to have a solution to both Eqs.\ (\ref
{eqp}) and (\ref{e1}) in the entire spacetime, or even in the entire vacuum
part of spacetime.

The nonexistence of a global solution $\Psi $ can be demonstrated from
another point of view. For a point particle in an otherwise-vacuum
spacetime, Barack and Ori \cite{BO} showed there is no global
radiation-gauge solution $h_{\alpha \beta }$ around the particle. As was
discussed in section \ref{sec1}, one can construct a solution $h_{\alpha
\beta }^{+}$ which is entirely regular at $r>r_{particle}$, but this
solution will necessarily have a singularity in the range $r<r_{particle}$,
along a line emanating from the particle. Alternatively one may construct a
solution $h_{\alpha \beta }^{-}$ which is regular in the entire domain $
r<r_{particle}$, but this solution will have a line singularity in the range 
$r>r_{particle}$.

Although it is not possible to construct a solution $\Psi $ valid in the
entire vacuum region, it is possible (for a point source) to construct a
solution which is valid everywhere throughout the vacuum region, except in a
set of zero measure. (In fact there are two such solutions, those denoted 
$\Psi ^{+}$ and $\Psi ^{-}$ in section \ref{sec1} above.) 
We shall now proceed to
construct such a solution. Specifically we shall describe the construction
of the solution $\Psi ^{+}$ (but the other solution $\Psi ^{-}$ 
may be constructed in a fully analogous manner). 
This construction is applicable in both cases of a point
source and an extended source [though in the latter case the domain in which 
$\Psi ^{+}$ violates the required equations is no longer of zero measure, as
we discuss in the next section.]

As in the previous sections, we shall consider here the function $\bar{\Psi}
^{+}$ sourced by a particular mode $\lambda m\omega $ of $\psi _{0}$. This
function takes the decomposed form (\ref{14}), and we need to construct the
radial function $\hat{R}_{-2}^{\lambda m\omega }$. In the vacuum region 
$r>r_{\max }$, $\hat{R}_{-2}^{\lambda m\omega }$ is just the solution
described in Section \ref{sec4}. This was shown to be a valid solution of
both Eqs.\ (\ref{11}) and (\ref{16}) (and this is the {\em only} valid
solution). Note that in this external region $\psi _{0}$ is made of outgoing
modes only, 
\begin{equation}
R_{+2}^{\lambda m\omega }(r)=A^{(out)}\,R_{+2}^{+}(r)
\,\,\,\,\,\,\,\,\,\,\,(r>r_{\max })\,,  \label{ext}
\end{equation}
where hereafter we denote the relevant basis functions for brevity as 
\[
R_{\pm 2}^{+}\equiv R_{\pm 2}^{\lambda m\omega (out)}\,\,,\,\,R_{\pm
2}^{-}\equiv R_{\pm 2}^{\lambda m\omega (down)}\,\,, 
\]
and similarly we use $C^{+}\equiv C^{(out)}$, $C^{-}\equiv C^{(down)}$.
(Only $R_{\pm 2}^{\lambda m\omega (out)}$ and $R_{\pm 2}^{\lambda m\omega
(down)}$ will be relevant here, because these are the two homogeneous basis
functions involved in the construction of the retarded Green's functions for
the Teukolsky variables $\psi _{\pm 2}$; see below). Therefore, in 
$r>r_{\max }$ the radial function of $\bar{
\Psi}^{+}$ (for a particular mode $\lambda m\omega $) is simply given by 
\begin{equation}
\hat{R}_{-2}^{\lambda m\omega
}(r)=C^{+}A^{(out)}R_{-2}^{+}(r)\,\,\,\,\,\,\,\,\,\,\,\,(r>r_{\max })\,.
\label{r>rmax}
\end{equation}

Consider next the extension of this solution into the range $r<r_{\max }$.
Here $R_{+2}^{\lambda m\omega }(r)$ is not a vacuum solution (it fails to be
a vacuum solution everywhere in $r_{\min }<r<r_{\max }$), and
we can no longer require $\hat{R}_{-2}^{\lambda m\omega }$ to satisfy both
Eqs.\ (\ref{11}) and (\ref{16}). \footnote{
For, extending $\hat{R}_{-2}^{\lambda m\omega }$ into $r<r_{\max }$ as a
solution of Eq.\ (\ref{16}) would automatically yield the external solution 
(\ref{r>rmax}) in this range too. But then $(D_{m\omega })^{4}[\hat{R}
_{-2}^{\lambda m\omega }]$ would necessarily be the analytically-extended
vacuum function $R_{+2}^{\lambda m\omega }$, which does not conform with the
actual, non-vacuum, function $R_{+2}^{\lambda m\omega }$ in $r_{\min
}<r<r_{\max }$.} We therefore choose to extend $\hat{R}_{-2}^{\lambda
m\omega }$ into $r<r_{\max }$ as a solution of Eq.\ (\ref{11}), and, for the
time being, forget about Eq.\ (\ref{16}). Nevertheless, in the next section
we shall show that this procedure yields a valid solution $\Psi ^{+}$, which
solves both required (time-domain) equations (\ref{eqp}) and (\ref{e1}),
even at $r<r_{\max }$ -- except in the domain $\Sigma ^{+}$ (which is of
zero measure for a point source).

To construct the solution of Eq.\ (\ref{11}) we proceed as follows. The
source term in this linear differential equation is $R_{+2}^{\lambda m\omega
}(r)$, the radial function of $\psi _{0}$. This function is in turn sourced
by the energy-momentum distribution in spacetime. It can thus be expressed
by means of the energy-momentum source term $T_{+2}^{\lambda m\omega }(r)$,
via the Green's-function method: 
\begin{equation}
R_{+2}^{\lambda m\omega }(r)=\int\limits_{r_{\min }}^{r_{\max
}}T_{+2}^{\lambda m\omega }(r^{\prime })G(r,r^{\prime })dr^{\prime }\,.
\label{green}
\end{equation}
The Green's function $G(r,r^{\prime })$ is constructed from the two vacuum
solutions admitting the desired asymptotic behavior, namely, outgoing waves
at large $r$ and down-going waves at the EH: 
\begin{eqnarray}
G(r,r^{\prime }) &=&A^{+}(r^{\prime })R_{+2}^{+}(r)\theta (r-r^{\prime }) 
\nonumber \\
&&+A^{-}(r^{\prime })R_{+2}^{-}(r)\,\theta (r^{\prime }-r)\,,
\label{standard}
\end{eqnarray}
where $\theta $ denotes the standard step function, namely $\theta (x)=0$
for $x<0$ and $\theta (x)=1$ for $x>0$. The functions $A^{+}(r^{\prime })$
and $A^{-}(r^{\prime })$ are given by 
\begin{equation}
A^{\pm }=R_{+2}^{\mp }/(\Delta W[R_{+2}^{-},R_{+2}^{+}])\,  \label{A0}
\end{equation}
(all quantities evaluated at $r^{\prime }$), where $W[R_{+2}^{-},R_{+2}^{+}]$
denotes the Wronskian of the two homogeneous solutions $R_{+2}^{\pm }$. Note
that $G$, viewed as a function of $r$, is continuous at $r=r^{\prime }$,
namely 
\begin{equation}
A^{+}(r^{\prime })R_{+2}^{+}(r^{\prime })=A^{-}(r^{\prime
})R_{+2}^{-}(r^{\prime })\,.  \label{AB}
\end{equation}
The coefficient $A^{(out)}$ in the external solution (\ref{ext}) is thus
given by 
\begin{equation}
A^{(out)}=\int\limits_{r_{\min }}^{r_{\max }}T_{+2}^{\lambda m\omega
}(r^{\prime })A^{+}(r^{\prime })dr^{\prime }\,.  \label{aout}
\end{equation}

We now wish to construct a ''Green-like function'' $H(r,r^{\prime })$ such
that the function $\hat{R}_{-2}^{\lambda m\omega }$ will be given by 
\begin{equation}
\hat{R}_{-2}^{\lambda m\omega }(r)=\int\limits_{r_{\min }}^{r_{\max
}}T_{+2}^{\lambda m\omega }(r^{\prime })H(r,r^{\prime })dr^{\prime }\,,
\label{Hgreen}
\end{equation}
in analogy with Eq.\ (\ref{green}). This will be a solution to Eq.\ (\ref{11})
if $H(r,r^{\prime })$ satisfies the equation 
\begin{equation}
(D_{m\omega })^{4}[H(r,r^{\prime })]=G(r,r^{\prime })\,  \label{HG}
\end{equation}
(in which the operator $D_{m\omega }$ differentiates with respect to $r$,
not $r^{\prime }$). Motivated by the above form of $G(r,r^{\prime })$, we
assume a function $H(r,r^{\prime })$ of a similar form, 
\begin{equation}
H(r,r^{\prime })=H^{+}(r,r^{\prime })\theta (r-r^{\prime
})+H^{-}(r,r^{\prime })\theta (r^{\prime }-r)\,,  \label{H+-}
\end{equation}
where $H^{+}(r,r^{\prime })$ and $H^{-}(r,r^{\prime })$ are smooth functions
of their arguments. Equation (\ref{HG}) is then satisfied if the following
two conditions hold: (i) the two functions $H^{\pm }(r,r^{\prime })$ satisfy 
\begin{equation}
(D_{m\omega })^{4}[H^{\pm }(r,r^{\prime })]=A^{\pm }(r^{\prime })R_{+2}^{\pm
}(r)\,,  \label{H+}
\end{equation}
and (ii) $H(r,r^{\prime })$ is continuous and differentiable four times
(with respect to $r)$ at $r=r^{\prime }$; In other words, the function 
\[
y(r,r^{\prime })\equiv H^{+}(r,r^{\prime })-H^{-}(r,r^{\prime })\,, 
\]
and its derivatives with respect to $r$ up to 4th order, vanish at 
$r=r^{\prime }$: 
\begin{equation}
\frac{\partial ^{n}y}{\partial r^{n}}(r=r^{\prime
})=0\,,\,\,\,\,\,\,\,\,\,\,n=0,...,4\,\,\,  \label{4der}
\end{equation}
(in which $\partial ^{0}y/\partial r^{0}\equiv y$ is to be understood).
Condition (i) guarantees the validity of Eq.\ (\ref{HG}) at $r>r^{\prime }$
and $r<r^{\prime }$ [the ''$+$'' and ''$-$'' cases in Eq.\ (\ref{H+}),
respectively]. Condition (ii) is required for the validity of Eq.\ (\ref{HG})
at $r=r^{\prime }$. [To see this, rewrite Eq.\ (\ref{H+-}) as 
\[
H(r,r^{\prime })=y(r,r^{\prime })\theta (r-r^{\prime })+H^{-}(r,r^{\prime
})\,, 
\]
and recall the continuity of $G$ at $r=r^{\prime }$.]

The form of the general solution $H^{\pm }(r,r^{\prime })$ to Eq.\ (\ref{H+})
is obvious from the analysis in sections \ref{sec4} and \ref{sec5} above. We
have a specific inhomogeneous solution $C^{\pm }A^{\pm }(r^{\prime
})R_{-2}^{\pm }(r)$, and the general homogeneous solution (\ref{general1})
(in which the coefficients $B_{i}$ are now allowed to be arbitrary functions
of $r^{\prime }$); Hence the most general solution is 
\begin{equation}
H^{\pm }(r,r^{\prime })=C^{\pm }A^{\pm }(r^{\prime })R_{-2}^{\pm
}(r)+e^{-i(mu-\omega r_{*})}\sum_{i=0}^{3}B_{i}^{\pm }(r^{\prime })r^{i}.
\label{H+gen}
\end{equation}
It should be noted that since this is the most general solution to Eq.\ (\ref
{11}), this form must be satisfied by both $\Psi ^{+}$ and $\Psi ^{-}$. The
difference between these two solutions should emerge from the choice of the
free functions $B_{i}^{\pm }$, which are to be determined by the boundary
conditions. As we are considering here the solution $\Psi ^{+}$, the radial
function $\hat{R}_{-2}^{\lambda m\omega }(r)$ must satisfy Eq.\ (\ref{r>rmax}) 
at $r>r_{\max }$. This is achieved by simply choosing $B_{i}^{+}(r^{\prime
})\equiv 0$ for all $i$, namely 
\begin{equation}
H^{+}(r,r^{\prime })=C^{+}A^{+}(r^{\prime })R_{-2}^{+}(r)\,.  \label{H+spec}
\end{equation}
The internal part $H^{-}(r,r^{\prime })$ has a more complicated form, 
\begin{equation}
H^{-}(r,r^{\prime })=C^{-}A^{-}(r^{\prime })R_{-2}^{-}(r)+e^{-i(mu-\omega
r_{*})}\sum_{i=0}^{3}B_{i}^{-}(r^{\prime })r^{i},  \label{H-gen}
\end{equation}
in which the four arbitrary functions $B_{i}^{-}(r^{\prime })$ are to be
determined by condition (ii) above, i.e. by matching to $H^{+}(r,r^{\prime
}) $ at $r=r^{\prime }$. The function $y(r,r^{\prime })$ is given by 
\begin{eqnarray*}
y(r,r^{\prime }) &=&C^{+}A^{+}(r^{\prime
})R_{-2}^{+}(r)-C^{-}A^{-}(r^{\prime })R_{-2}^{-}(r) \\
&&-e^{-i(mu-\omega r_{*})}\sum_{i=0}^{3}B_{i}^{-}(r^{\prime })r^{i}\,,
\end{eqnarray*}
and we must impose Eq.\ (\ref{4der}). This might look problematic at first
glance, because apparently the latter equation imposes five requirements on
the four arbitrary functions $B_{i}^{-}(r^{\prime })$. However, one of these
requirements is automatically satisfied. To see this, it will be convenient
to rewrite Eq.\ (\ref{4der}) as 
\[
(D_{m\omega
})^{n}[y]=0\,,\,\,\,\,\,\,\,\,\,\,\,\,\,\,n=0,...,4\,\,\,\,\,\,\,\,\,\,\,\,
\,\,\,\,\,(r=r^{\prime })\, 
\]
(where $(D_{m\omega })^{0}[y]\equiv y$ is to be understood). Considering the
case $n=4$, the operator $(D_{m\omega })^{4}\ $annihilates the last term in
the above expression for $y(r,r^{\prime })$, and we have 
\[
(D_{m\omega })^{4}[y]=A^{+}(r^{\prime })R_{+2}^{+}(r)-A^{-}(r^{\prime
})R_{+2}^{-}(r)\,, 
\]
which vanishes by virtue of Eq.\ (\ref{AB}). We can therefore re-express
condition (ii) as 
\begin{equation}
(D_{m\omega
})^{n}[y]=0\,,\,\,\,\,\,\,\,\,\,\,\,\,\,\,n=0,...,3\,\,\,\,\,\,\,\,\,\,\,\,
\,\,\,\,\,(r=r^{\prime })\,.  \label{3D}
\end{equation}
The four arbitrary functions $B_{i}^{-}(r^{\prime })$ should thus be
determined from the four conditions involved in this equation. In Appendix A
we show that these four functions may be expressed as 
\begin{eqnarray*}
B_{i}^{-}(r^{\prime }) &=&C^{+}A^{+}(r^{\prime })\left[ f_{i}(r^{\prime
})R_{-2}^{+}(r^{\prime })+g_{i}(r^{\prime })\frac{d}{dr^{\prime }}
R_{-2}^{+}(r^{\prime })\right] \\
&&-C^{-}A^{-}(r^{\prime })\left[ f_{i}(r^{\prime })R_{-2}^{-}(r^{\prime
})+g_{i}(r^{\prime })\frac{d}{dr^{\prime }}R_{-2}^{-}(r^{\prime })\right]
\end{eqnarray*}
where $f_{i}(r^{\prime })$ and $g_{i}(r^{\prime })$ are certain functions of 
$r^{\prime }$ explicitly specified therein.

\subsection{Further simplification of the solution}

The procedure described so far for the construction of $\hat{R}
_{-2}^{\lambda m\omega }(r)$ requires the two $s=-2$ homogeneous basis
functions $R_{-2}^{\pm }$, and also the two $s=+2$ homogeneous functions $
R_{+2}^{\pm }$. The latter functions are required for the determination of
$A^{\pm }(r^{\prime })$ (and their derivatives are involved in
the Wronskian $W[R_{+2}^{-},R_{+2}^{+}]$). However, in
principle $\psi _{0}$ can be determined from $\psi _{4}$ (and vice versa),
and this implies that $R_{+2}^{\pm }$ may be determined from $R_{-2}^{\pm }$.
In Appendix B we undertake this goal and re-express $A^{\pm }(r^{\prime })$
in terms of the functions $R_{-2}^{\pm }$ and their first-order derivatives.
We find 
\begin{eqnarray}
A^{\pm }(r^{\prime }) &=&\left( pC^{\pm }W[R_{-2}^{-},R_{-2}^{+}]\right)
^{-1}\times  \nonumber \\
&&\left[ \bar{A}(r^{\prime })\frac{d}{dr^{\prime }}R_{-2}^{\mp }(r^{\prime
})+\bar{B}(r^{\prime })R_{-2}^{\mp }(r^{\prime })\right] \,,  \label{ApAm}
\end{eqnarray}
where $p$ is the parameter defined in Eq.\ (\ref{p}), $\bar{A}$ and $\bar{B}$
are functions specified in Eq.\ (\ref{ABfinal}), and 
\begin{equation}
W[R_{-2}^{-},R_{-2}^{+}]=const\cdot \Delta (r^{\prime })\,  \label{wronsk}
\end{equation}
is the Wronskian of the two basis solutions $R_{-2}^{\pm }$ (evaluated at 
$r^{\prime }$).

In the above construction of $H(r,r^{\prime })$, 
$A^{\pm }$ and $C^{\pm }$ only appear through their products $A^{+}C^{+}$ and 
$A^{-}C^{-}$. We therefore define 
\[
a^{\pm }(r^{\prime })\equiv C^{\pm }A^{\pm }(r^{\prime })\,, 
\]
and obtain 
\begin{eqnarray}
a^{\pm }(r^{\prime }) &=&\left( pW[R_{-2}^{-},R_{-2}^{+}]\right) ^{-1}\times
\nonumber \\
&&\left[ \bar{A}(r^{\prime })\frac{d}{dr^{\prime }}R_{-2}^{\mp }(r^{\prime
})+\bar{B}(r^{\prime })R_{-2}^{\mp }(r^{\prime })\right] \,.
\end{eqnarray}
The functions $H^{\pm }(r,r^{\prime })$ and $B_{n}^{-}(r^{\prime })$ can now
be re-expressed as 
\begin{equation}
H^{+}(r,r^{\prime })=a^{+}(r^{\prime })R_{-2}^{+}(r)\,,
\end{equation}
\begin{equation}
H^{-}(r,r^{\prime })=a^{-}(r^{\prime })R_{-2}^{-}(r)+e^{-i(mu-\omega
r_{*})}\sum_{i=0}^{3}B_{i}^{-}(r^{\prime })r^{i}\,,
\end{equation}
and 
\begin{eqnarray*}
B_{i}^{-}(r^{\prime }) &=&a^{+}(r^{\prime })\left[ f_{i}(r^{\prime
})R_{-2}^{+}(r^{\prime })+g_{i}(r^{\prime })\frac{d}{dr^{\prime }}
R_{-2}^{+}(r^{\prime })\right] \\
&&-a^{-}(r^{\prime })\left[ f_{i}(r^{\prime })R_{-2}^{-}(r^{\prime
})+g_{i}(r^{\prime })\frac{d}{dr^{\prime }}R_{-2}^{-}(r^{\prime })\right] \,.
\end{eqnarray*}

Note that when expressed in this form the function $H(r,r^{\prime })$ ---
and hence also $\hat{R}_{-2}^{\lambda m\omega }(r)$ --- is invariant to a
rescaling of $R_{-2}^{+}$ or $R_{-2}^{-}$ by constants. Therefore there is
no need to require here a specific normalization for these basis solutions.

\section{Domain of validity of the constructed solution}

\label{sec7}

In the case of perturbations produced by matter sources, we must
carefully examine in what parts of spacetime the construction of $\Psi $ and 
$h_{\alpha \beta }$ is valid. First, the Chrzanowski's construction requires
that the potential $\Psi $ satisfies both equations (\ref{eqp}) 
and (\ref{e1}). 
These equations are mutually consistent in vacuum, but are generally
inconsistent in the presence of matter. The matter source 
therefore leads to violation
of at least one of these equations. This violation
occurs not only in the region occupied by the matter, but also in
certain vacuum parts of the spacetime.\footnote{
To verify this, note that in the above construction of $\Psi ^{+}$, the
individual-mode radial function $\hat{R}_{-2}^{\lambda m\omega }(r)$
violates the radial Teukolsky equation in the entire domain $r<r_{\max }$
--- which in particular includes the vacuum domain $r<r_{\min }$. This
violation follows from the non-vanishing of the coefficients $B_{i}^{-}$.
This does not necessarily mean that the time-domain Teukolsky equation (\ref
{eqp}) is violated everywhere in $r<r_{\max }$ (in fact it does not, as we
show below); but it does indicate the existence of a domain of violation
that extends at any $r$ value in $r<r_{\max }$.}

At this stage it will be conceptually simpler to assume that the massive
object that creates the perturbation has a finite size and a regular
energy-momentum distribution $T_{\mu \nu }(x^{\alpha })$. (The case of a
point mass will then follow in a trivial manner.) Let us define $\Sigma ^{+}$
to be the collection of all points P in spacetime which have the following
property: The null geodesic $\xi $ (a member of the outgoing principal null
congruence) passing through P intersects matter (more precisely,
nonvanishing source term $T_{+2}$) on its approach from P towards future
null infinity. The collection of all other points of
(the part $r>r_{+}$ of) spacetime is denoted 
$\hat{\Sigma }^{+}$. In an analogous manner, we define $\Sigma ^{-}$ to be
the collection of all points P for which $\xi $ intersects matter on its
approach from P towards the EH, and $\hat{\Sigma }^{-}$ is the rest 
of (the part $r>r_{+}$ of)
spacetime. In a more pictorial language, imagine that light rays propagate
along all geodesics $\xi $ of the outgoing principal null congruence. Then 
$\Sigma ^{-}$ is the portion of spacetime ''shadowed'' by the matter-energy
distribution $T_{+2}(x^{\alpha })$, and $\hat{\Sigma }^{-}$ is the
non-shadowed part. Similarly, $\Sigma ^{+}$ is the ''past-shadowed'' part,
i.e. the part of spacetime that would be shadowed if the light rays were
propagating along the null congruence from future to past (and from large $r$
towards the EH), and $\hat{\Sigma }^{+}$ is the rest of $r>r_{+}$. By
definition, $\hat{\Sigma }^{+}$ and $\hat{\Sigma }^{-}$ are pure vacuum
domains. Note that $\hat{\Sigma }^{+}$ contains the entirely-vacuum domain 
$r>r_{\max }$, and generally 
it also extends through $r<r_{\max }$ into the other
entirely-vacuum domain $r<r_{\min }$ --- though the latter domain is not
entirely contained in $\hat{\Sigma }^{+}$. (Similarly, $\hat{\Sigma }^{-}$
contains the entire domain $r<r_{\min }$, and generally 
extends through $r>r_{\min }$ into $r>r_{\max }$).
Also, at the point-like limit $\Sigma ^{+}$ and $\Sigma ^{-}$ each
degenerates to a $(1+1)$-dimensional surface, that emerges out of the
particle's worldline in either the past or future (i.e. inward or outward)
direction of $\xi $. Hence in the point-like limit $\hat{\Sigma }^{+}$ or 
$\hat{\Sigma }^{-}$ cover the entire spacetime except a set of measure zero.
On the other hand, when the object is extended, $\Sigma ^{+}$ and 
$\Sigma^{-} $ are four-dimensional sets.

We shall now argue that the above constructed potential $\Psi ^{+}$ --- and
the MP $h_{\alpha \beta }^{+}$ constructed from it via Eq.\ (\ref{MP}) ---
are valid in the entire domain $\hat{\Sigma }^{+}$. We shall first establish
this for the potential $\Psi ^{+}$, namely, we shall show that Eqs.\ (\ref
{eqp}) and (\ref{e1}) are satisfied throughout $\hat{\Sigma }^{+}$. Then we
shall show that the MP solution $h_{\alpha \beta }^{+}$ constructed from 
$\Psi ^{+}$ is valid throughout $\hat{\Sigma }^{+}$ too. (The same arguments
apply to the validity of $\Psi ^{-}$ and $h_{\alpha \beta }^{-}$ throughout 
$\hat{\Sigma }^{-}$.)

\subsection{Validity of the constructed potential $\Psi ^{+}$}

In the range $r>r_{\max }$, the above construction of the ''Green-like function'' 
$H(r,r^{\prime })$ ensures that Eq.\ (\ref{r>rmax}) is satisfied by $\hat{R}
_{-2}^{\lambda m\omega }(r)$, and hence also Eqs.\ (\ref{11}) and (\ref{16}).
In the time domain this implies that $\Psi ^{+}$ satisfies Eqs.\ (\ref{eqp})
and (\ref{e1}) throughout $r>r_{\max }$.

In the range $r<r_{\max }$ we have constructed $\hat{R}_{-2}^{\lambda
m\omega }(r)$ such that it satisfies Eq.\ (\ref{11}) for all modes, hence in
the time domain compliance with Eq.\ (\ref{e1}) is guaranteed. But Eq.\ (\ref
{16}) is violated throughout $r<r_{\max }$ [note that the homogeneous
solution (\ref{general1}) violates Eq.\ (\ref{16})]. We shall now employ
analyticity considerations to demonstrate that $\Psi ^{+}$ does satisfy the 
{\em time-domain} Teukolsky equation (\ref{eqp}) throughout $\hat{\Sigma }
^{+}$.

Each mode $\lambda m\omega $ of $\bar{\Psi} ^{+}$ is analytic throughout 
$r>r_{\max }$ by construction [cf. Eq.\ (\ref{r>rmax})].
Assuming convergence of the mode sum at 
$r>r_{\max }$, we may assume that $\bar{\Psi} ^{+}$ itself is analytic at 
$r>r_{\max }$ too.\footnote{
For our argument to hold it is sufficient that the mode sum converges
throughout some range $r>r_{\max }^{\prime }\geq r_{\max }$, or even
throughout some open interval of $r$ values located 
somewhere at $r>r_{\max }$.} Now, in the above
construction we extended $\bar{\Psi} ^{+}$ 
into the entire domain $r<r_{\max }$ as
a solution of Eq.\ (\ref{e1}). This implies that $\bar{\Psi} ^{+}$ is analytic
throughout $\hat{\Sigma }^{+}$, as we now show.

Equation (\ref{e1}) is an ordinary differential equation along the null
geodesics $\xi $, which we may write as 
\begin{equation}
\frac{d^{4}\bar{\Psi}^{+}(\gamma ;\xi )}{d\gamma ^{4}}=\psi _{0}(\gamma ;\xi
)\,\,.  \label{ord1}
\end{equation}
We write its general solution explicitly (in a recursive manner) as $\bar{
\Psi}^{+}(\gamma ;\xi )\equiv \Phi _{1}(\gamma ;\xi )$, with 
\begin{equation}
\Phi _{n}(\gamma ;\xi )=c_{n}(\xi )+\int\limits_{r_{0}}^{\gamma }\Phi
_{n+1}(\gamma ^{\prime };\xi )d\gamma ^{\prime
}\,\,\,\,\,\,\,\,\,\,\,\,\,\,\,\,\,\,n=1...4\,\,,  \label{solution}
\end{equation}
where $\Phi _{5}\equiv \psi _{0}$. Here $c_{i}(\xi )$ ($i=1...4$) are four
arbitrary functions of the three variables $\theta _{0},t_{0},\varphi _{0}$.
[Recall that the geodesics $\xi $ are parametrized by the three quantities 
$\theta _{0},t_{0},\varphi _{0}$, defined through Eq.\ (\ref{parameters}),
that take constant values along the geodesic. Also recall that we have set 
$\gamma =r.$] We take the lower integration limit to be, say, $r_{0}=2r_{\max
}$. The transformation from the coordinates ($t,r,\theta ,\varphi $) to 
($\gamma ,\theta _{0},t_{0},\varphi _{0}$) can be read off Eq.\ (\ref
{parameters}), and it is manifestly analytic everywhere in $r>r_{+}$.
Therefore, the analyticity of $\bar{\Psi}^{+}$ in the domain $r>r_{\max }$
implies it is analytic in ($\gamma ,\theta _{0},t_{0},\varphi _{0}$) as
well. This in turn implies that all four functions $c_{i}(\xi )\equiv
c_{i}(\theta _{0},t_{0},\varphi _{0})$ are analytic in ($\theta
_{0},t_{0},\varphi _{0}$). Now, the function $\psi _{0}$ is presumably
analytic everywhere in the vacuum region.\footnote{
In the point-like case $\psi _{0}$ is irregular at the particle's location.
In the case of a smooth extended source, $\psi _{0}$ will fail to be
analytic at the boundary of the region occupied by matter. But in both cases
we may assume that $\psi _{0}$ is analytic throughout the
vacuum region.} When the solution (\ref{solution}) is restricted to the
domain $\hat{\Sigma }^{+}$, we observe that only vacuum points $(\gamma
^{\prime };\xi )$ are encountered in the integration, hence $\psi _{0}(\gamma
^{\prime };\xi )$ is analytic. This immediately implies that $\bar{\Psi}^{+}$
given in Eq.\ (\ref{solution}) is analytic in ($\gamma ;\theta
_{0},t_{0},\varphi _{0}$) throughout $\hat{\Sigma }^{+}$, and hence also in 
($t,r,\theta ,\varphi $).

From the analyticity of $\bar{\Psi}^{+}$ 
(which implies the analyticity of $\Psi^{+}$) it follows that
$W_{-2}\left[ \Psi ^{+}\right] $ is analytic too. The vanishing of the latter
at $r>r_{\max }$ therefore implies its vanishing throughout $\hat{\Sigma }^{+}$.
We have thus established the compliance of $\Psi ^{+}$ with Eqs.\ (\ref
{eqp}) and (\ref{e1}) throughout $\hat{\Sigma }^{+}$.

It is easy to see why this argument fails at $\Sigma ^{+}$: The function 
$\psi _{0}$ fails to be analytic at the point particle, or --- in the case of
an extended object --- at the boundary of the matter distribution. As a
consequence, along each null geodesic $\xi $ intersecting the source, $\bar{\Psi}
^{+}$ will be analytic only up to the intersection point. Then $\bar{\Psi}^{+}$
will usually be non-analytic at the boundary of $\Sigma ^{+}$.
Therefore we cannot expect $\Psi ^{+}$ to satisfy Eq.\ (\ref{eqp}) in $\Sigma
^{+}$. The violation of the corresponding frequency-domain equation (\ref{16}) 
throughout $r<r_{\max }$ indicates that Eq.\ (\ref{eqp}) is indeed violated
somewhere in $\Sigma ^{+}$ (and for any value of $r$ in this range).

Finally we note that the compliance of $\Psi ^{+}$ with the ``angular equation'' 
(i.e. Eq.\ (2.7) in Ref.\ \cite{LW}) throughout $\hat{\Sigma }^{+}$ 
may be deduced by exactly the same analyticity argument.

\subsection{Validity of the constructed metric perturbation}

Our goal here is to establish the validity of $h_{\alpha \beta }^{+}$
(constructed from $\Psi ^{+}$ via the Chrzanowski's method) throughout $\hat{
\Sigma }^{+}$, despite the presence of matter in spacetime. By ''validity''
we mean that (i) $h_{\alpha \beta }^{+}$ satisfies the linearized vacuum
Einstein equations, and (ii) the $s=+2$ Weyl scalar constructed from it
coincides with the original field $\psi _{0}$. To this end we use
analyticity considerations, similar to those used above 
for analyzing the validity of $\Psi ^{+}$. 
Here we shall briefly sketch these considerations. \cite{wald1}

Consider first the validity of $h_{\alpha \beta }^{+}$ in $r>r_{\max }$. To
this end, expand $\Psi ^{+}$ (and $\psi _{0}$) into modes. For a particular
mode $\lambda m\omega $, extend the $r>r_{\max }$ vacuum solution
analytically into $r>r_{\max }$. This extended solution represents a pure
vacuum perturbation. Chrzanowski's construction may now
be applied to it, yielding the MP solution $h_{\alpha \beta
}^{+\lambda m\omega }= \Pi \left[ \Psi ^{+\lambda m\omega }\right] $
for the mode under consideration. Upon summation over
the modes, we obtain a valid MP solution 
$h_{\alpha \beta }^{+}=\Pi [\Psi^{+}]$ 
in the range $r>r_{\max }$.

Next consider the validity of the solution $h_{\alpha \beta }^{+}\equiv \Pi
\left[ \Psi ^{+}\right] $ in the part $r<r_{\max }$ of $\hat{\Sigma }^{+}$.
From the analyticity of $\Psi ^{+}$ throughout $\hat{\Sigma }^{+}$
(established above) it follows that $h_{\alpha \beta }^{+}$ is also analytic
in this range. Recall also the analyticity of $\psi _{0}$ throughout $\hat{
\Sigma }^{+}$. The above criteria (i,ii) for the validity of a MP solution 
$h_{\alpha \beta }$ are both formulated in terms of analytic differential
operators acting on $h_{\alpha \beta }$. From the validity of these
criteria in the range $r>r_{\max }$ it now follows that they must hold
throughout $\hat{\Sigma }^{+}$.

\section{Summary of main results: Gravitational perturbations}

\label{sec8}

Here we briefly summarize our procedure for constructing the potential $\Psi 
$, for gravitational perturbations. We use the decomposition 
\[
\bar{\Psi}=\sum_{\lambda m\omega }\hat{R}_{-2}^{\lambda m\omega
}S_{+2}^{\lambda m\omega }(\theta )e^{i(m\varphi -\omega t)}\,, 
\]
and our goal is to construct the radial functions $\hat{R}_{-2}^{\lambda
m\omega }$. We shall now summarize this construction in the two different
cases: (i) pure gravitational waves, and (ii) perturbations with sources.

\subsection{Pure gravitational waves}

In this cases we assume that $\psi _{0}$ is given. This field is decomposed
into modes too, 
\[
\psi _{0}=\sum_{\lambda m\omega }R_{+2}^{\lambda m\omega }(r)S_{+2}^{\lambda
m\omega }(\theta )e^{i(m\varphi -\omega t)}\,. 
\]
For each mode $\lambda m\omega $, $R_{+2}^{\lambda m\omega }(r)$ is a
solution of the vacuum radial Teukolsky equation, and we assume this
function is provided as a linear combination of two basis solutions. Two
sets of convenient basis solutions are: (i) the {\em large-}$r${\em \ set} ,
in which the basis solutions for $R_{+2}^{\lambda m\omega }$ and 
$R_{-2}^{\lambda m\omega }$ are given in Eqs.\ (\ref{in},\ref{out}), and (ii)
the {\em EH set}, in which the basis solutions for $R_{\pm 2}^{\lambda
m\omega }$ are given in Eqs.\ (\ref{hor+},\ref{hor-}).

Assume now that the information about $\psi _{0}$ is given in terms of any
two of the above four $s=+2$ basis functions, namely

\[
R_{+2}^{\lambda m\omega }(r)=A^{(a)}R_{+2}^{\lambda m\omega
(a)}(r)+A^{(b)}R_{+2}^{\lambda m\omega (b)}(r)\,, 
\]
and the coefficients $A^{(a)}$ and $A^{(b)}$ are provided for each mode 
$\lambda m\omega $. (Here ''$a$'' and ''$b$'' denote either the large-$r$
basis solutions, or the horizon basis solutions, or any combination of these
two sets, e.g. ''$a$''$=$''$out$'' and ''$b$''$=$''$down$''). Then, the
corresponding radial functions $\hat{R}_{-2}^{\lambda m\omega }$ of $\bar{
\Psi}$ are simply given by 
\[
\hat{R}_{-2}^{\lambda m\omega }(r)=C^{(a)}A^{(a)}R_{-2}^{\lambda m\omega
(a)}(r)+C^{(b)}A^{(b)}R_{-2}^{\lambda m\omega (b)}(r)\,. 
\]
The four coefficients $C^{(in)}$, $C^{(out)}$, $C^{(down)}$ and $C^{(up)}$
are specified in Eqs.\ (\ref{cf},\ref{cdown},\ref{cup}).

\subsection{Gravitational perturbations produced by sources}

Here we consider the case in which the perturbation is produced by a
distribution of matter-energy. This may be either a point-like particle, or
an extended object. In both cases we assume that we are given the radial
energy-momentum 
source function $T_{+2}^{\lambda m\omega }(r)$ for each mode [this is the
source term in the $s=+2$ radial Teukolsky equation (\ref{Tr})]. For
simplicity we assume here that the matter source is restricted to the range 
$r_{\min }\leq r\leq r_{\max }$ (but this assumption may be relaxed -- at
least partially -- as we discuss in section \ref{sec10}).

Then $\hat{R}_{-2}^{\lambda m\omega }(r)$ is given by 
\begin{equation}
\hat{R}_{-2}^{\lambda m\omega }(r)=\int\limits_{r_{\min }}^{r_{\max
}}T_{+2}^{\lambda m\omega }(r^{\prime })H(r,r^{\prime })dr^{\prime }\,\,,
\end{equation}
where 
\begin{equation}
H(r,r^{\prime })=H^{+}(r,r^{\prime })\theta (r-r^{\prime
})+H^{-}(r,r^{\prime })\theta (r^{\prime }-r)\,,
\end{equation}
and $H^{\pm }(r,r^{\prime })$ are two smooth functions. We construct these
functions from the two $s=-2$ homogeneous radial solutions 
$R_{-2}^{\lambda m\omega (out)}\equiv R_{-2}^{+}$ and 
$R_{-2}^{\lambda m\omega (down)}\equiv R_{-2}^{-}$, defined by their
asymptotic behavior

\[
R_{-2}^{+}(r)\propto r^{3}e^{i\omega
r_{*}}\,\,\,\,\,\,\,\,\,\,\,\,\,(r\to \infty ) 
\]
and 
\[
R_{-2}^{-}(r)\propto \Delta
^{2}e^{-ikr*}\,\,\,\,\,\,\,\,\,\,\,\,(r_{*}\to -\infty )\,, 
\]
where $k=\omega -ma/(2Mr_{+})$. (The $s=+2$ basis solutions are not required
here. Also we do not require here a specific normalization for $R_{-2}^{+}$
and $R_{-2}^{-}$.) We find 
\begin{equation}
H^{+}(r,r^{\prime })=a^{+}(r^{\prime })R_{-2}^{+}(r)\,
\end{equation}
and 
\begin{equation}
H^{-}(r,r^{\prime })=a^{-}(r^{\prime })R_{-2}^{-}(r)+e^{-i(mu-\omega
r_{*})}\sum_{i=0}^{3}B_{i}^{-}(r^{\prime })r^{i}\,,
\end{equation}
where $u$ and $r_{*}$ are defined in Eqs.\ (\ref{u},\ref{r*d}), respectively,
and 
\begin{eqnarray}
a^{\pm }(r^{\prime }) &=&\left( pW[R_{-2}^{-},R_{-2}^{+}]\right) ^{-1}\times
\nonumber \\
&&\left[ \bar{A}(r^{\prime })\frac{d}{dr^{\prime }}R_{-2}^{\mp }(r^{\prime
})+\bar{B}(r^{\prime })R_{-2}^{\mp }(r^{\prime })\right] \,.
\end{eqnarray}
Here $p$ is a parameter given in Eq.\ (\ref{p}), and 
$W[R_{-2}^{-},R_{-2}^{+}] $ is the Wronskian of the two basis functions
(which is proportional to $\Delta $), evaluated at $r^{\prime }$. The
functions $\bar{A}(r^{\prime }),\bar{B}(r^{\prime })$ are specified in
Appendix B. The four functions $B_{i}^{-}(r^{\prime })$ are given by 
\begin{eqnarray*}
B_{i}^{-}(r^{\prime }) &=&a^{+}(r^{\prime })\left[ f_{i}(r^{\prime
})R_{-2}^{+}(r^{\prime })+g_{i}(r^{\prime })\frac{d}{dr^{\prime }}
R_{-2}^{+}(r^{\prime })\right] \\
&&-a^{-}(r^{\prime })\left[ f_{i}(r^{\prime })R_{-2}^{-}(r^{\prime
})+g_{i}(r^{\prime })\frac{d}{dr^{\prime }}R_{-2}^{-}(r^{\prime })\right] \,,
\end{eqnarray*}
where $f_{i}$ and $g_{i}$ are functions of $r^{\prime }$ specified in
Appendix A.

The above construction yields the radial functions $\hat{R}_{-2}^{\lambda
m\omega }(r)$ for the solution $\Psi ^{+}$ that is valid (and regular)
through $\hat{\Sigma }^{+}$. This domain includes the entire range 
$r>r_{\max }$, but not all points of $r<r_{\max }$. The other solution $\Psi
^{-}$ that is valid through $\hat{\Sigma }^{-}$ (i.e. everywhere in 
$r<r_{\min }$ but not at all points of $r>r_{\min }$) may be constructed in a
fully analogous manner. The only difference is in the functions $H^{\pm
}(r,r^{\prime })$, which now take the forms 
\begin{eqnarray*}
H^{+}(r,r^{\prime }) &=&a^{+}(r^{\prime })R_{-2}^{+}(r) \\
&&-e^{-i(mu-\omega r_{*})}\sum_{i=0}^{3}B_{i}^{-}(r^{\prime
})r^{i}\,\,\,\,\,\,\,\,\,\,\,\,\,\,\,\,\,\,(\Psi ^{-})\,,\,
\end{eqnarray*}
\[
H^{-}(r,r^{\prime })=a^{-}(r^{\prime
})R_{-2}^{-}(r)\,\,\,\,\,\,\,\,\,\,\,\,\,\,\,\,\,\,\,\,\,\,\,\,\,\,\,(\Psi
^{-})\,. 
\]

\section{Summary of main results: Electromagnetic perturbations}

\label{sec9}

The electromagnetic case is treated in full analogy with the gravitational
case. Here, again, the four-potential $A_{\alpha }$ is constructed in
Chrzanowski's method, by applying a certain differential operator $\Pi _{EM}$
to a potential $\Psi _{EM}$: 
\begin{equation}
A_{\alpha }=\Pi _{EM}\left[ \Psi _{EM}\right] \,.
\end{equation}
Throughout this section we shall denote the electromagnetic potential $\Psi
_{EM}$ as $\Psi $ for brevity. This potential satisfies equations analogous
to Eqs.\ (\ref{eqp}) and (\ref{e1}): 
\begin{equation}
W_{-1}\left[ \Psi \right] =0\,
\end{equation}
and 
\begin{equation}
\varphi _{0}=-D^{2}[\bar{\Psi}]\,,  \label{DDEM}
\end{equation}
where $\varphi _{0}$ is the $s=+1$ Weyl scalar, and $W_{-1}$ is the $s=-1$
case of the differential operator (\ref{meq}). [See e.g. \cite{wald}, in
which $\Psi $ is denoted ''$\varphi _{E}$''. The last equation is the
reduction of Eq.\ (15) therein to the Kerr case.] We use the decomposition 
\[
\bar{\Psi}=\sum_{\lambda m\omega }\hat{R}_{-1}^{\lambda m\omega
}S_{+1}^{\lambda m\omega }(\theta )e^{i(m\varphi -\omega t)}\,, 
\]
and our goal is to construct the radial functions $\hat{R}_{-1}^{\lambda
m\omega }$. Again, we shall construct this function first in the case of
pure electromagnetic waves, and then for perturbations with sources. Here we
shall summarize the results; The main steps in the derivations are presented
in Appendix C.

\subsection{Pure electromagnetic waves}

In this cases we assume that $\varphi _{0}$ is given. This field is
decomposed into modes as 
\[
\varphi _{0}=\sum_{\lambda m\omega }R_{+1}^{\lambda m\omega
}(r)S_{+1}^{\lambda m\omega }(\theta )e^{i(m\varphi -\omega t)}\,. 
\]
For each mode $\lambda m\omega $, $R_{+1}^{\lambda m\omega }(r)$ is a
solution of the vacuum radial Teukolsky equation, and we assume this
function is provided as a linear combination of two basis solutions. The two
sets of convenient basis solutions are: (i) the {\em large-}$r${\em \ set} , 
\begin{equation}
R_{+1}^{\lambda m\omega (in)}\cong e^{-i\omega
r_{*}}/r\,\,,\,\,\,\,\,\,R_{+1}^{\lambda m\omega (out)}\cong e^{i\omega
r_{*}}/r^{3}\,\,,
\end{equation}
\begin{equation}
R_{-1}^{\lambda m\omega (in)}\cong e^{-i\omega
r_{*}}/r\,\,,\,\,\,\,\,\,R_{-1}^{\lambda m\omega (out)}\cong re^{i\omega
r_{*}}{}\,;
\end{equation}
and (ii) the {\em EH set}, 
\begin{equation}
R_{+1}^{\lambda m\omega (down)}\cong \Delta
^{-1}e^{-ikr*}\,\,,\,\,\,\,\,\,R_{+1}^{\lambda m\omega (up)}\cong
e^{ikr*}\,,
\end{equation}
\begin{equation}
R_{-1}^{\lambda m\omega (down)}\cong \Delta
e^{-ikr*}\,\,,\,\,\,\,\,\,R_{-1}^{\lambda m\omega (up)}\cong e^{ikr*}\,,
\end{equation}
where $k=\omega -ma/(2Mr_{+})$.

Assume now that $\varphi _{0}$ is given in terms of any two of the above
four basis functions for $R_{+1}^{\lambda m\omega }$, namely

\begin{equation}
R_{+1}^{\lambda m\omega }(r)=A^{(a)}R_{+1}^{\lambda m\omega
(a)}(r)+A^{(b)}R_{+1}^{\lambda m\omega (b)}(r)\,,
\end{equation}
and the coefficients $A^{(a)}$ and $A^{(b)}$ are provided for each mode $
\lambda m\omega $. (Here, again, ''$a$'' and ''$b$'' denote two of the above
four basis functions, e.g. ''$a$''$=$''$out$'' and ''$b$''$=$''$down$''.)
Then, the corresponding radial functions $\hat{R}_{-1}^{\lambda m\omega }$
of $\bar{\Psi}$ are given by 
\begin{equation}
\hat{R}_{-1}^{\lambda m\omega }(r)=C^{(a)}A^{(a)}R_{-1}^{\lambda m\omega
(a)}(r)+C^{(b)}A^{(b)}R_{-1}^{\lambda m\omega (b)}(r)\,.
\end{equation}
The four coefficients $C^{(in)}$, $C^{(out)}$, $C^{(down)}$,$C^{(up)}$ take
now the values 
\begin{equation}
C^{(in)}=1/(4\omega ^{2})\,,\,C^{(out)}=4\omega
^{2}/p\,\,\,\,\,\,\,\,\,\,\,\,\text{(EM)}\,\,,
\end{equation}
\begin{equation}
C^{(up)}=\bar{Q}/p\,,\,C^{(down)}=1/Q\,\,\,\,\,\,\,\,\,\,\,\,\,\text{(EM)}\,,
\end{equation}
where in the electromagnetic case we have 
\[
p=\lambda ^{2}-4\alpha ^{2}\omega ^{2}=\lambda ^{2}+4a\omega \left
(m-a\omega \right) \,\,\,\,\,\,\,\,\,\,\text{(EM)} 
\]
and 
\begin{equation}
Q=w(w+iq)\,,\,\,\,\,\,\,\,\,\,\text{(EM)}
\end{equation}
and, recall, 
\[
w=4kMr_{+}\,\,,\,\,\,q=r_{+}-r_{-}=2(M^{2}-a^{2})^{1/2}\,. 
\]

\subsection{Electromagnetic perturbations produced by sources}

Here we consider the case in which the perturbation is produced by charges
and/or currents (e.g. a point charge or an extended charged object orbiting
the BH). We assume that we are given the radial electromagnetic source
function $T_{+1}^{\lambda m\omega }(r)$ for each mode [this is the source
term in the $s=+1$ analog of the radial Teukolsky equation (\ref{Tr})]. As
before we assume for simplicity that the source is restricted to the range 
$r_{\min }\leq r\leq r_{\max }$.

The radial function $\hat{R}_{-1}^{\lambda m\omega }(r)$ then takes the form 
\begin{equation}
\hat{R}_{-1}^{\lambda m\omega }(r)=\int\limits_{r_{\min }}^{r_{\max
}}T_{+1}^{\lambda m\omega }(r^{\prime })H(r,r^{\prime })dr^{\prime }\,\,,
\end{equation}
where 
\begin{equation}
H(r,r^{\prime })=H^{+}(r,r^{\prime })\theta (r-r^{\prime
})+H^{-}(r,r^{\prime })\theta (r^{\prime }-r)\,,
\end{equation}
and $H^{\pm }(r,r^{\prime })$ are two smooth functions. We construct these
functions from the two $s=-1$ homogeneous radial solutions 
$R_{-1}^{\lambda m\omega (out)}\equiv R_{-1}^{+}$ and 
$R_{-1}^{\lambda m\omega (down)}\equiv R_{-1}^{-}$, defined by their
asymptotic behavior

\[
R_{-1}^{+}(r)\propto re^{i\omega
r_{*}}\,\,\,\,\,\,\,\,\,\,\,\,\,(r\to \infty ) 
\]
and 
\[
R_{-1}^{-}(r)\propto \Delta
e^{-ikr*}\,\,\,\,\,\,\,\,\,\,\,\,(r_{*}\to -\infty )\,. 
\]
(Here, again, we do not require a specific normalization for 
$R_{-1}^{\pm }$). We find 
\begin{equation}
H^{+}(r,r^{\prime })=a^{+}(r^{\prime })R_{-1}^{+}(r)\,
\end{equation}
and 
\begin{equation}
H^{-}(r,r^{\prime })=a^{-}(r^{\prime })R_{-1}^{-}(r)+e^{-i(mu-\omega
r_{*})}\sum_{i=0}^{1}B_{i}^{-}(r^{\prime })r^{i}\,,
\end{equation}
where $u$ and $r_{*}$ are defined in Eqs.\ (\ref{u},\ref{r*d}), respectively,
and 
\begin{eqnarray}
a^{\pm }(r^{\prime }) &=&-\left( pW[R_{-1}^{-},R_{-1}^{+}]\right)
^{-1}\,\times  \nonumber \\
&&\left[ \bar{A}(r^{\prime })\frac{d}{dr^{\prime }}R_{-1}^{\mp }(r^{\prime
})+\bar{B}(r^{\prime })R_{-1}^{\mp }(r^{\prime })\right] \,.
\end{eqnarray}
Here $W[R_{-1}^{-},R_{-1}^{+}]=const$ is the Wronskian of the two basis
functions $R_{-1}^{\pm }$, and 
\[
\bar{A}(r^{\prime })=2iK\,\,,\,\,\bar{B}(r^{\prime })=\lambda +2i\omega
r^{\prime }-2K^{2}/\Delta \,\,\,\,\,\,\,\,\,\,\,\,\,\,\text{(EM)}\, 
\]
[with all quantities evaluated at $r^{\prime }$, e.g. $K=am-(r^{\prime
2}+a^{2})$]. The two functions $B_{i}^{-}(r^{\prime })$ are given by 
\begin{eqnarray*}
B_{i}^{-}(r^{\prime }) &=&a^{+}(r^{\prime })\left[ f_{i}(r^{\prime
})R_{-1}^{+}(r^{\prime })+g_{i}(r^{\prime })\frac{d}{dr^{\prime }}
R_{-1}^{+}(r^{\prime })\right] \\
&&-a^{-}(r^{\prime })\left[ f_{i}(r^{\prime })R_{-1}^{-}(r^{\prime
})+g_{i}(r^{\prime })\frac{d}{dr^{\prime }}R_{-1}^{-}(r^{\prime })\right]
\end{eqnarray*}
(for $i=0,1$), where the functions $f_{i}(r^{\prime }),g_{i}(r^{\prime })$
are

\begin{eqnarray*}
f_{0}(r^{\prime }) &=&\left( 1-iKr^{\prime }/\Delta \right) e^{i(mu-\omega
r_{*})}\,\,,\,\, \\
g_{0}(r^{\prime }) &=&-r^{\prime }e^{i(mu-\omega r_{*})}\,,
\end{eqnarray*}

\[
f_{1}(r^{\prime })=(i/\Delta )Ke^{i(mu-\omega
r_{*})}\,\,\,,\,\,\,\,g_{1}(r^{\prime })=e^{i(mu-\omega
r_{*})}\,\,\,\,\,\,\,\, 
\]
(again, with $u,r_{*},K,\Delta $ all evaluated at $r^{\prime }$).

The above construction yields the radial functions $\hat{R}_{-1}^{\lambda
m\omega }(r)$ for the solution $\Psi ^{+}$ that is valid (and regular)
throughout $\hat{\Sigma }^{+}$. The other solution $\Psi ^{-}$ that is valid
throughout $\hat{\Sigma }^{-}$ is constructed in a fully analogous manner.
The only difference is in the functions $H^{\pm }(r,r^{\prime })$, which now
take the forms 
\begin{eqnarray*}
H^{+}(r,r^{\prime }) &=&a^{+}(r^{\prime })R_{-1}^{+}(r) \\
&&-e^{-i(mu-\omega r_{*})}\sum_{i=0}^{1}B_{i}^{-}(r^{\prime
})r^{i}\,\,\,\,\,\,\,\,\,\,\,\,\,\,\,(\Psi ^{-})\,,
\end{eqnarray*}
\[
H^{-}(r,r^{\prime })=a^{-}(r^{\prime
})R_{-1}^{-}(r)\,\,\,\,\,\,\,\,\,\,\,\,\,\,\,\,\,\,\,\,\,\,\,\,\,\,\,\,\,\,
(\Psi ^{-})\,. 
\]

\section{Discussion}

\label{sec10}

Although in most of this paper we referred explicitly to gravitational
perturbations, the same construction applies to the electromagnetic case as
well, as outlined in section \ref{sec9}. In particular, the domains of
validity are the same in both cases: $\hat{\Sigma }^{+}$ for $\Psi ^{+}$
(and for $h_{\alpha \beta }^{+}$ or $A_{\alpha }^{+}$ derived from the
latter), and $\hat{\Sigma }^{-}$ for $\Psi ^{-}$ (and for $h_{\alpha \beta
}^{-}$ or $A_{\alpha }^{-}$).

Also, although we have explicitly considered the ingoing radiation
gauge throughout this paper, an analogous construction may be applied to the 
{\em outgoing} radiation gauge. In this latter gauge, too, there are two
solutions, $\Psi _{ORG}^{+}$ and $\Psi _{ORG}^{-}$ (and correspondingly 
$h_{ORG}^{+},A_{ORG}^{+}$ and $h_{ORG}^{-},A_{ORG}^{-}$), which are valid in
the domains $\hat{\Sigma }_{ORG}^{+}$ and $\hat{\Sigma }_{ORG}^{-}$ (but
invalid in $\Sigma _{ORG}^{+}$ or $\Sigma _{ORG}^{-}$), respectively. The
two domains $\hat{\Sigma }_{ORG}^{\pm }$ are completely analogous to $\hat{
\Sigma }^{\pm }\equiv \hat{\Sigma }_{IRG}^{\pm }$, except that they are
defined with respect to the {\em ingoing} rather than outgoing principal
null congruence. In the rest of this discussion, too, we shall refer
explicitly to the ingoing gauge, but the same remarks will be applicable to
the outgoing gauge as well.

Consider the case of a point particle. Our analysis shows there does not
exist a single solution for the radiation-gauge $h_{\alpha \beta }$ or 
$A_{\alpha }$ that is regular in the entire off-worldline neighborhood of the
particle. Instead, the solution $\Psi ^{+}$ (and correspondingly $h_{\alpha
\beta }^{+},A_{\alpha }^{+}$) has a line singularity along the outgoing null
geodesic $\xi $ emanating from the particle towards the past and smaller $r$. 
Similarly, $\Psi ^{-}$ (and correspondingly $h_{\alpha \beta
}^{-},A_{\alpha }^{-}$) has a line singularity along the null geodesic $\xi $
emanating from the particle towards the future and larger $r$. The
inevitability of such a line singularity in the radiation-gauge MP was
previously demonstrated in Ref.\ \cite{BO} based on independent arguments.
(The existence of ingoing radiation-gauge solutions other than $h_{\alpha
\beta }^{\pm },A_{\alpha }^{\pm }$, which admit a line singularity in a
different direction, not tangent to $\xi $, has not been explored yet.)

The unavoidable occurrence of a line singularity in the radiation-gauge
fields $h_{\alpha \beta },$ $A_{\alpha }$ is obviously an inconvenient
property. Nevertheless it does not pose a too serious obstacle (at least in
some important applications). We must recall that this singularity is after
all a gauge artifact, which may in principle be removed by an appropriate
gauge transformation. Therefore, whenever the local values of $h_{\alpha
\beta }$ or $A_{\alpha }$ are required for the calculation of any local
gauge-invariant quantity, the solutions $h_{\alpha \beta }^{+},$ $A_{\alpha
}^{+}$ and/or $h_{\alpha \beta }^{-},$ $A_{\alpha }^{-}$ may be used
regardless of the line singularity.

An important application which requires the knowledge of $h_{\alpha \beta }$
or $A_{\alpha }$ is the radiation-reaction problem for a point mass or point
charge. Generically the full analysis of this phenomenon requires the
calculation of the local self force acting on the particle. The
electromagnetic self force is gauge-invariant. The situation in the
gravitational problem is more delicate, because the gravitational self force
is a gauge-dependent entity. Nevertheless, within the context of the
adiabatic approximation, the orbit-integrated change (induced by the self
force) in any of the orbit's constants of motion is gauge-invariant. One
thus may use any gauge to calculate the self force, and hence the rate of
change of the constants of motion. Consider the calculation of the self
force according to the Mino-Sasaki-Tanaka \cite{MST} formulation. Then the
self force is the limit of the ''tail-force'' field at the particle's
location. This limit may be taken from any desired direction. Two especially
convenient directions are the ingoing and outgoing radial directions (so far
the mode-sum method \cite{BMNOS} has been fully developed for these radial
directions only.) To this end, one may use the solution $h_{\alpha \beta
}^{+}$ or $A_{\alpha }^{+}$ when calculating the self force from the radial
direction $r>r_{particle}$, and the solution $h_{\alpha \beta }^{-}$ or $
A_{\alpha }^{-}$ for calculating the self force from $r<r_{particle}$. In both
cases the line singularity is not encountered. \footnote{
Recall, however, that in the gravitational case there is another difficulty
associated with the radiation gauge: The leading-order asymptotic behavior
of the MP, on approaching the particle's location from a generic direction,
differs from that of the harmonic-gauge MP, making this an
''irregular gauge'' in the terminology of Ref.\ \cite{BO}. This kind of
irregularity (which is unrelated to the line singularity on 
$\Sigma ^{\pm }$) also occurs in e.g. the Regge-Wheeler gauge in the
Schwarzschild case. This difficulty may in principle be overcome by
transforming to an ''intermediate gauge'', as outlined in Ref.\ \cite{BO}.}

In the case of a smooth extended source, $\Sigma ^{+}$ (or $\Sigma ^{-}$)
becomes a four-dimensional set. In this case $\Psi ^{+}$ does not develope
an irregularity at $\Sigma ^{+}$; however, the equation (\ref{eqp}) is
violated there. This {\em suggests} that the quantity $h_{\alpha \beta }^{+}$
(constructed from $\Psi ^{+}$ by applying the differential operator $\Pi $)
will not be valid at $\Sigma ^{+}$, even in its vacuum part (namely, it will
fail to satisfy the vacuum Einstein equation, and/or to reproduce the
original Teukolsky field $\psi _{0}$); But this still needs be verified.

In the above construction we have assumed that the particle's worldline or
the extended source is restricted to a range $r_{\min }\leq r\leq r_{\max }$.
This assumption was made primarily for conceptual clarity, as it allows us
to discuss the behavior of e.g. $\Psi ^{+}$ in the two vacuum regions, 
$r>r_{\max }$ and $r<r_{\min }$; but it can be relaxed at least partially, as
we now discuss.

Consider, first, the situation in which the source is restricted to the
range $r\leq r_{\max }$ with no minimal value $r_{\min }$. \footnote{
In the case of a point particle, this situation may be realized by a
''fine-tuned'' geodesic that asymptotes to
an unstable circular orbit in the far past,
but falls into the BH in the future. We prefer not to consider here bound
geodesics emerging out of the white-hole horizon, to avoid 
the conceptual complications associated with the latter's causal properties.}
Then the construction of $\Psi ^{+}$ follows just as prescribed above,
without any difficulties. The construction of $\Psi ^{-}$ in this case may
formally be carried out as above; However, the proof given in section \ref
{sec7} for the validity of $\Psi ^{-}$ throughout $\hat{\Sigma }^{-}$ fails
in this case: This proof (when applied to $\Psi ^{-}$ rather than $\Psi ^{+}
$) starts from the trivial observation that (provided that the source is
restricted to $r\geq r_{\min }$) Eq.\ (\ref{eqp}) is satisfied by $\Psi ^{-}$
throughout $r<r_{\min }$. Then this feature is analytically extended to the
entire domain $\hat{\Sigma }^{-}$. In the present case (i.e. no $r_{\min }$)
this proof is inapplicable even at its starting point. 
It therefore still needs be verified whether in this case 
the so constructed solution $\Psi ^{-}$ is valid in $\hat{\Sigma }^{-}$.

In the analogous case, in which the source extends from infinity to some 
$r_{\min }$, the situation is basically similar, though technically it is
slightly more involved. Consider for example an unbounded orbit that arrives
from infinity and scatters off the BH back to infinity. Here, the solution 
$\Psi ^{-}$ can in principle be constructed as above, but the solution $\Psi
^{+}$ is not guaranteed to hold (for the reason explained just above). In
this case, however, due to the slow decay at large $r$ of the potential
term in the radial Teukolsky equation, the standard integral solution (\ref
{green},\ref{standard}) for $\psi _{0}$ diverges. 
One then has to use another Green's
function \cite{poisson} for the construction of $\psi _{0}$, and this may
modify the function $H(r,r^{\prime })$. We shall not elaborate on this case
here.

Finally we note that there are a few types of special modes which
require special treatment. First, for the stationary modes $\omega =0$,
the large-$r$ basis solutions $R_{\pm 2}^{\lambda m\omega (in,out)}$
constructed in section \ref{sec4} must be replaced by some other ones, and
the same holds for the corresponding constants $C^{(in)}$ and $C^{(out)}$. 
Second,
for ''marginally-superradiant'' modes $k=0$, the EH basis solutions $R_{\pm
2}^{\lambda m\omega (up,down)}$ and the corresponding constants $
C^{(up,down)}$ are to be modified. It appears likely, 
though, that in both cases
the inhomogeneous solution described in e.g. section \ref{sec8} remains valid,
provided that one substitutes 
the appropriate basis functions $R_{-2}^{\pm }(r)$ (i.e.
those satisfying the correct boundary conditions at large $r$ or at the EH).
Other cases which require special attention are the so-called 
``$l=0,1$ modes'' (the ``$l=0$ mode'' in the electromagnetic case). 
These are the perturbation modes for which the Teukolsky variables 
$\psi _{0}$ and $\psi _{4}$ vanish identically, while $\psi _{2}$ is 
nonvanishing.
The extension of this construction to include the ``$l=0,1$ modes'', 
as well as all other $\omega =0$ modes, is now underway.

\section*{Acknowledgment}

I would like to thank B. Whiting and R. Wald for interesting discussions and
for bringing to my attention several delicate aspects of the problem. 
This research was supported by The Israel Science Foundation 
(grant no. 74/02-11.1), 
the Fund for the Promotion of Research at the Technion, 
and the Technion V.P.R. Fund.

\appendix

\label{A}%%%%    Appendix A    %%%
%
%%%%   \section{}      %%%

\section{} 

We rewrite Eq.\ (\ref{3D}) as 
\begin{eqnarray}
&&(D_{m\omega })^{n}[e^{-i(mu-\omega
r_{*})}\sum_{i=0}^{3}B_{i}^{-}(r^{\prime })r^{i}]  \nonumber \\
&=&C^{+}A^{+}(r^{\prime })(D_{m\omega })^{n}[R_{-2}^{+}(r)]  \nonumber \\
&&-C^{-}A^{-}(r^{\prime })(D_{m\omega })^{n}[R_{-2}^{-}(r)]\,  \label{41}
\end{eqnarray}
(applied at $r=r^{\prime }$ and for $n=0,...,3$). It will be convenient to
rewrite the polynomial $\Sigma _{i}B_{i}^{-}(r^{\prime })r^{i}$ as $\Sigma
_{i}\hat{B}_{i}(r^{\prime })(r-r^{\prime })^{i}$. By virtue of Eq.\ (\ref
{Diff}), the left-hand side of the last equation reads 
\[
e^{-i(mu-\omega r_{*})}\sum_{i=0}^{3}\hat{B}_{i}(r^{\prime })\frac{d^{n}}{
dr^{n}}(r-r^{\prime })^{i}=e^{-i(mu-\omega r_{*})}n!\hat{B}_{n}(r^{\prime
})\,. 
\]
Evaluating Eq.\ (\ref{41}) at $r=r^{\prime }$ then implies (for $n=0,...,3$) 
\begin{eqnarray}
e^{-i(mu-\omega r_{*})}n!\hat{B}_{n}(r^{\prime }) &=&C^{+}A^{+}(r^{\prime
})(D_{m\omega })^{n}[R_{-2}^{+}]  \nonumber \\
&&-C^{-}A^{-}(r^{\prime })(D_{m\omega })^{n}[R_{-2}^{-}]  \label{42}
\end{eqnarray}
(where $D_{m\omega }^{n}[R_{-2}^{\pm }]$ is to be evaluated at $r=r^{\prime
} $). The operator $D_{m\omega }$ is given by 
\[
D_{m\omega }=\frac{d}{dr}+(i/\Delta )K\,. 
\]
Using the radial Teukolsky equation (\ref{vacuum}) we can express any
derivative of an $s=-2$ vacuum radial function $R_{-2}^{\lambda m\omega }$
as a linear combination of $R_{-2}^{\lambda m\omega }$ and 
$(d/dr)R_{-2}^{\lambda m\omega }$ (and consequently we can express any power
of $D_{m\omega }$ as a linear combination of $R_{-2}^{\lambda m\omega }$ and 
$D_{m\omega }[R_{-2}^{\lambda m\omega }]$). As a consequence, when applied
to any homogeneous solutions $R_{-2}^{\lambda m\omega }$ (and in particular $
R_{-2}^{\pm }$), we have the following operator identities: \cite{DDD} 
\begin{eqnarray*}
(D_{m\omega })^{2} &=&(2/\Delta )(iK+r-M)D_{m\omega }+\frac{\lambda
+6i\omega r}{\Delta } \\
&=&(2/\Delta )(iK+r-M)\frac{d}{dr} \\
&&+\Delta ^{-2}[2iK(iK+r-M)+\Delta (\lambda +6i\omega r)]
\end{eqnarray*}
and

\begin{eqnarray*}
(D_{m\omega })^{3} &=&\Delta ^{-2}[4iK(iK+r-M)+(\lambda +2-2i\omega r)\Delta
]D_{m\omega } \\
&&+\Delta ^{-2}[2iK(\lambda +6i\omega r)+6i\omega \Delta ]\, \\
&=&\Delta ^{-2}[4iK(iK+r-M)+(\lambda +2-2i\omega r)\Delta ]\frac{d}{dr} \\
&&+\Delta ^{-3}\{iK[4iK(iK+r-M) \\
&&+(\lambda +2-2i\omega r)\Delta ] \\
&&+\Delta [2iK(\lambda +6i\omega r)+6i\omega \Delta ]\,\}\,.
\end{eqnarray*}
Thus, we may write Eq.\ (\ref{42}) as 
\begin{eqnarray*}
\hat{B}_{n}(r^{\prime }) &=&C^{+}A^{+}(r^{\prime })\left[ \hat{f}
_{n}(r^{\prime })R_{-2}^{+}(r^{\prime })+\hat{g}_{n}(r^{\prime })\frac{d}{
dr^{\prime }}R_{-2}^{+}(r^{\prime })\right] \\
&&-C^{-}A^{-}(r^{\prime })\left[ \hat{f}_{n}(r^{\prime
})R_{-2}^{-}(r^{\prime })+\hat{g}_{n}(r^{\prime })\frac{d}{dr^{\prime }}
R_{-2}^{-}(r^{\prime })\right] \,
\end{eqnarray*}
(for $n=0,...,3$), where the functions $\hat{f}_{n},\hat{g}_{n}$ are given
by 
\[
\hat{f}_{0}(r)=e^{i(mu-\omega r_{*})}\,,\,\,\hat{g}_{0}(r)=0\,, 
\]
\[
\hat{f}_{1}(r)=(i/\Delta )Ke^{i(mu-\omega r_{*})}\,,\,\,\hat{g}
_{1}(r)=e^{i(mu-\omega r_{*})}\,, 
\]
\[
\hat{f}_{2}(r)=(2\Delta ^{2})^{-1}[2iK(iK+r-M)+\Delta (\lambda +6i\omega
r)]e^{i(mu-\omega r_{*})}\,,\,\, 
\]
\[
\hat{g}_{2}(r)=\Delta ^{-1}(iK+r-M)e^{i(mu-\omega r_{*})}\,, 
\]
\begin{eqnarray*}
\hat{f}_{3}(r) &=&(6\Delta ^{3})^{-1}\{iK[4iK(iK+r-M) \\
&&+(\lambda +2-2i\omega r)\Delta ] \\
&&+\Delta [2iK(\lambda +6i\omega r)+6i\omega \Delta ]\,\}e^{i(mu-\omega
r_{*})}\,,
\end{eqnarray*}
\begin{eqnarray*}
\hat{g}_{3}(r) &=&(6\Delta ^{2})^{-1}[4iK(iK+r-M) \\
&&+(\lambda +2-2i\omega r)\Delta ]e^{i(mu-\omega r_{*})}\,.
\end{eqnarray*}

Once the coefficients $\hat{B}_{i}(r^{\prime })$ are determined, the
original coefficients $B_{n}^{-}(r^{\prime })$ may be constructed through 
\[
\sum_{i=0}^{3}B_{i}^{-}(r^{\prime })r^{i}=\sum_{i=0}^{3}\hat{B}
_{i}(r^{\prime })(r-r^{\prime })^{i}\,, 
\]
which yields 
\begin{eqnarray*}
B_{0}^{-} &=&\hat{B}_{0}-\hat{B}_{1}r^{\prime }+\hat{B}_{2}r^{\prime 2}-\hat{
B}_{3}r^{\prime 3}\,,\,\,\,\, \\
B_{1}^{-} &=&\hat{B}_{1}-2\hat{B}_{2}r^{\prime }+3\hat{B}_{3}r^{\prime 2}\,,
\\
B_{2}^{-} &=&\hat{B}_{2}-3\hat{B}_{3}r^{\prime
}\,\,\,\,,\,\,\,\,\,\,\,B_{3}^{-}=\hat{B}_{3}\,\,.
\end{eqnarray*}
This allows us to express the functions $B_{n}^{-}(r^{\prime })$ as

\begin{eqnarray*}
B_{n}^{-}(r^{\prime }) &=&C^{+}A^{+}(r^{\prime })\left[ f_{n}(r^{\prime
})R_{-2}^{+}(r^{\prime })+g_{n}(r^{\prime })\frac{d}{dr^{\prime }}
R_{-2}^{+}(r^{\prime })\right] \\
&&-C^{-}A^{-}(r^{\prime })\left[ f_{n}(r^{\prime })R_{-2}^{-}(r^{\prime
})+g_{n}(r^{\prime })\frac{d}{dr^{\prime }}R_{-2}^{-}(r^{\prime })\right] \,,
\end{eqnarray*}
with the functions $f_{n}(r^{\prime }),g_{n}(r^{\prime })$ given by 
\begin{eqnarray*}
f_{0} &=&\hat{f}_{0}-\hat{f}_{1}r^{\prime }+\hat{f}_{2}r^{\prime 2}-\hat{f}
_{3}r^{\prime 3}\,,\,\,\,\,f_{1}=\hat{f}_{1}-2\hat{f}_{2}r^{\prime }+3\hat{f}
_{3}r^{\prime 2}\,, \\
f_{2} &=&\hat{f}_{2}-3\hat{f}_{3}r^{\prime }\,,\,\,\,\,f_{3}=\hat{f}_{3}\,,
\end{eqnarray*}
and similarly 
\begin{eqnarray*}
g_{0} &=&\hat{g}_{0}-\hat{g}_{1}r^{\prime }+\hat{g}_{2}r^{\prime 2}-\hat{g}
_{3}r^{\prime 3}\,,\,\,\,\,g_{1}=\hat{g}_{1}-2\hat{g}_{2}r^{\prime }+3\hat{g}
_{3}r^{\prime 2}\,, \\
g_{2} &=&\hat{g}_{2}-3\hat{g}_{3}r^{\prime }\,,\,\,\,g_{3}=\hat{g}_{3}\,\,
\end{eqnarray*}
(with all functions  $\hat{g}_{n}$, $\hat{f}_{n}$ evaluated at 
$r^{\prime }$ rather than $r$).

\label{B}%%%%    Appendix B   %%%
%
%%%%   \section{}      %%%

\section{} 

Our goal is to construct the basis functions $R_{+2}^{\pm }$ from the
corresponding functions $R_{-2}^{\pm }$. To simplify the notation,
throughout this appendix we shall view $R_{\pm 2}^{\pm }$ and all other
''radial'' variables as functions or $r$, not $r^{\prime }$. When
implementing the result (\ref{finalB}) back in section \ref{sec6}, one
should simply substitute $r\to r^{\prime }$.

The analysis in section \ref{sec4}, Eqs.\ (\ref{cout},\ref{hdown}), implies 
\[
H[R_{+2}^{\pm }]=C^{\pm }R_{-2}^{\pm }\,, 
\]
which [since $H$ is the inverse of the operator $(D_{m\omega })^{4}$] is
equivalent to

\[
R_{+2}^{\pm }=C^{\pm }(D_{m\omega })^{4}[R_{-2}^{\pm }]\,. 
\]
With the aid of the radial Teukolsky equation, the operator $(D_{m\omega
})^{4}$ acting on any vacuum solution $R_{-2}^{\lambda m\omega }$ may be
expressed in terms of $R_{-2}^{\lambda m\omega }$ and its first-order
derivative. Chandrasekhar derived the formula [see Eq.\ (49CH), where
throughout this Appendix ''CH'' refers to equations in chapter 9 of Ref.\ 
\cite{chandra}] 
\begin{equation}
(D_{m\omega })^{4}[R_{-2}^{\lambda m\omega }]=(A_{0}/\Delta ^{3})D_{m\omega
}[R_{-2}^{\lambda m\omega }]+(B_{0}/\Delta ^{3})R_{-2}^{\lambda m\omega }\,,
\label{DDDD}
\end{equation}
where 
\begin{eqnarray}
A_{0} &=&-8iK[K^{2}+(r-M)^{2}]  \nonumber \\
&&+[4iK(\lambda +2)-8i\omega r(r-m)]\Delta +8i\omega \Delta ^{2}\,,
\end{eqnarray}
\begin{eqnarray}
B_{0} &=&[(\lambda +2-2i\omega r)(\lambda +6i\omega r)+12i\omega
(iK-r+M)]\Delta  \nonumber \\
&&+4iK(iK-r+M)(\lambda +6i\omega r)\,.
\end{eqnarray}
This yields 
\begin{equation}
R_{+2}^{\pm }=(C^{\pm }/\Delta ^{3})\left( A_{0}D_{m\omega }[R_{-2}^{\pm
}]+B_{0}R_{-2}^{\pm }\right) \,.  \label{R+2}
\end{equation}

We also need to express the determinant $W[R_{+2}^{-},R_{+2}^{+}]$ in terms
of $R_{-2}^{\pm }$. We find it useful to express this determinant as 
\begin{eqnarray*}
W[R_{+2}^{-},R_{+2}^{+}] &\equiv
&R_{+2}^{-}R_{+2,r}^{+}-R_{+2}^{+}R_{+2,r}^{-} \\
&=&R_{+2}^{-}{\cal D}_{-1}^{\dagger }[R_{+2,r}^{+}]-R_{+2,r}^{+}{\cal D}
_{-1}^{\dagger }[R_{+2}^{-}]\,,
\end{eqnarray*}
where ${\cal D}_{-1}^{\dagger }\equiv \partial _{r}-(iK/\Delta
)-2(r-M)/\Delta $, which allows us to make use of Eq.\ (50CH). Writing 
\[
{\cal D}_{-1}^{\dagger }(A_{0}D_{m\omega }+B_{0})=A_{1}D_{m\omega }+B_{1}\, 
\]
[with $A_{1},B_{1}$ specified in Eq.\ (51CH)], we obtain 
\[
W[R_{+2}^{-},R_{+2}^{+}]=C^{+}C^{-}\Delta
^{-6}(B_{0}A_{1}-A_{0}B_{1})W[R_{-2}^{-},R_{-2}^{+}]\,. 
\]
A straightforward calculation yields 
\[
B_{0}A_{1}-A_{0}B_{1}=p\Delta ^{2}\,, 
\]
leading to 
\begin{equation}
W[R_{+2}^{-},R_{+2}^{+}]=C^{+}C^{-}p\Delta ^{-4}W[R_{-2}^{-},R_{-2}^{+}]\,.
\label{wrons2}
\end{equation}
[Note the consistency of this result with the general expression for the
Wronskian of the Teukolsky equation: For any $s$, and any pair of
independent solutions $R_{s}^{a},R_{s}^{b}$, 
\[
W[R_{s}^{a},R_{s}^{b}]=const\cdot \Delta ^{-s-1}\,. 
\]
Hence $W[R_{+2}^{-},R_{+2}^{+}]=const\cdot \Delta ^{-3}$ and $
W[R_{-2}^{-},R_{-2}^{+}]=const\cdot \Delta $, in agreement with Eq.\ (\ref
{wrons2}).] Combining this result with Eqs.\ (\ref{R+2}) and (\ref{A0}), we
obtain 
\[
A^{\pm }=\left( C^{\pm }pW[R_{-2}^{-},R_{-2}^{+}]\right) ^{-1}\left
(A_{0}D_{m\omega }[R_{-2}^{\mp }]+B_{0}R_{-2}^{\mp }\right) \,. 
\]
This yields 
\begin{equation}
C^{\pm }A^{\pm }=\left( pW[R_{-2}^{-},R_{-2}^{+}]\right) ^{-1}\left[ \bar{A}
\frac{d}{dr}R_{-2}^{\mp }+\bar{B}R_{-2}^{\mp }\right] \,,  \label{finalB}
\end{equation}
with 
\begin{equation}
\bar{A}=A_{0}\,,\,\bar{B}=B_{0}+(iK/\Delta )A_{0}\,.  \label{ABfinal}
\end{equation}

\label{C}%%%%    Appendix C    %%%
%
%%%%   \section{}      %%%

\section{} 

For a vacuum mode $\lambda m\omega $ of electromagnetic perturbations, the
radial function $\hat{R}_{-1}^{\lambda m\omega }$ must satisfy the two
equations 
\begin{equation}
P_{-1}^{\lambda m\omega }[\hat{R}_{-1}^{\lambda m\omega }(r)]=0\,
\end{equation}
and 
\begin{equation}
R_{+1}^{\lambda m\omega }(r)=-(D_{m\omega })^{2}[\hat{R}_{-1}^{\lambda
m\omega }(r)]\,.  \label{DD}
\end{equation}
The general solution for these two equations is 
\begin{equation}
\hat{R}_{-1}^{\lambda m\omega }\equiv -p^{-1}\Delta (D_{m\omega }^{\dagger
})^{2}\Delta \left[ R_{+1}^{\lambda m\omega }(r)\right] \,,
\end{equation}
where 
\[
p=\lambda ^{2}-4\alpha ^{2}\omega ^{2}=\lambda ^{2}+4a\omega \left
(m-a\omega \right) \,\,\,\,\,\,\,\,\,\,\,\,\text{(EM)}\,. 
\]

Considering the four asymptotic basis solutions $R_{\pm 1}^{\lambda m\omega
(in,out,up,down)}$ specified in section \ref{sec9}, the corresponding
parameters $C^{(in)}$, $C^{(out)}$, $C^{(down)}$,$C^{(up)}$ are easily
calculated just as in the gravitational case. One finds 
\[
C^{(in)}=1/(4\omega ^{2})\,,\,C^{(out)}=4\omega ^{2}/p\, \, \, \, \, \, \,
\, \, \, \, \, \text{(EM)}\,\, 
\]
\[
C^{(up)}=Q/p\,,\,C^{(down)}=1/Q\,\, \, \, \, \, \, \, \, \, \, \, \, \, \,
\, \text{(EM)}\, . 
\]

The general solution to the homogeneous part of Eq.\ (\ref{DD}), namely 
$(D_{m\omega })^{2}[\hat{R}_{-1}^{\lambda m\omega }]=0$, is easily
constructed: 
\begin{equation}
\hat{R}_{-1}^{\lambda m\omega }(r)=e^{-i(mu-\omega
r_{*})}\sum_{i=0}^{1}B_{i}r^{i}\,.
\end{equation}

Consider next the case of inhomogeneous electromagnetic perturbations. The
general solution for the radial function of $\varphi _{0}$ may be expressed
as 
\begin{equation}
R_{+1}^{\lambda m\omega }(r)=\int\limits_{r_{\min }}^{r_{\max
}}T_{+1}^{\lambda m\omega }(r^{\prime })G(r,r^{\prime })dr^{\prime }\,.
\end{equation}
The Green's function is 
\[
G(r,r^{\prime })=A^{+}(r^{\prime })R_{+1}^{+}(r)\theta (r-r^{\prime
})+A^{-}(r^{\prime })R_{+1}^{-}(r)\,\theta (r^{\prime }-r)\,, 
\]
where 
\begin{equation}
A^{\pm }=R_{+1}^{\mp }/(\Delta W[R_{+1}^{-},R_{+1}^{+}])\,  \label{AEM}
\end{equation}
(evaluated at $r^{\prime }$). Then $\hat{R}_{-1}^{\lambda m\omega }(r)$ is
given by 
\begin{equation}
\hat{R}_{-1}^{\lambda m\omega }(r)=\int\limits_{r_{\min }}^{r_{\max
}}T_{+1}^{\lambda m\omega }(r^{\prime })H(r,r^{\prime })dr^{\prime }\,,
\end{equation}
where $H(r,r^{\prime })$ satisfies

\begin{equation}
(D_{m\omega })^{2}\left[ H(r,r^{\prime })\right] =-G(r,r^{\prime })\,
\end{equation}
(and the appropriate boundary conditions at $r>r_{\max }$). We find 
$H(r,r^{\prime })$ to be of the form 
\begin{equation}
H(r,r^{\prime })=H^{+}(r,r^{\prime })\theta (r-r^{\prime
})+H^{-}(r,r^{\prime })\theta (r^{\prime }-r)\,,
\end{equation}
with

\begin{equation}
H^{+}(r,r^{\prime })=C^{+}A^{+}(r^{\prime })R_{-1}^{+}(r)\,
\end{equation}
and 
\begin{equation}
H^{-}(r,r^{\prime })=C^{-}A^{-}(r^{\prime })R_{-1}^{-}(r)+e^{-i(mu-\omega
r_{*})}\sum_{i=0}^{1}B_{i}^{-}(r^{\prime })r^{i}\,.
\end{equation}
The two functions $B_{i}^{-}(r^{\prime })$ are determined by regularity
conditions at $r=r^{\prime }$, which yield 
\begin{eqnarray*}
B_{i}^{-}(r^{\prime }) &=&C^{+}A^{+}(r^{\prime })\left[ f_{i}(r^{\prime
})R_{-1}^{+}(r^{\prime })+g_{i}(r^{\prime })\frac{d}{dr^{\prime }}
R_{-1}^{+}(r^{\prime })\right] \\
&&-C^{-}A^{-}(r^{\prime })\left[ f_{i}(r^{\prime })R_{-1}^{-}(r^{\prime
})+g_{i}(r^{\prime })\frac{d}{dr^{\prime }}R_{-1}^{-}(r^{\prime })\right] .
\end{eqnarray*}
In full analogy with the gravitational case (see Appendix A) we find 
\[
f_{0}=\hat{f}_{0}-\hat{f}_{1}r^{\prime }\,,\,\,\,\,f_{1}=\hat{f}
_{1}\,\,,\,\,g_{0}=\hat{g}_{0}-\hat{g}_{1}r^{\prime }\,,\,\,\,\,g_{1}=\hat{g}
_{1}\,, 
\]
and 
\[
\hat{f}_{0}(r)=e^{i(mu-\omega r_{*})}\,,\,\,\hat{g}_{0}(r)=0\,, 
\]
\[
\hat{f}_{1}(r)=(i/\Delta )Ke^{i(mu-\omega r_{*})}\,,\,\,\hat{g}
_{1}(r)=e^{i(mu-\omega r_{*})}\,, 
\]
yielding the functions $f_{i},g_{i}$ specified in section \ref{sec9}.

Next we express $R_{+1}^{\mp }$ in terms of $R_{-1}^{\mp }$, using 
\begin{equation}
R_{+1}^{\pm }=-C^{\pm }(D_{m\omega })^{2}[R_{-1}^{\pm }]\,.  \label{71}
\end{equation}
The vacuum radial Teukolsky equation for the $s=-1$ radial function may be
expressed as \cite{chandra} 
\[
D_{m\omega }^{\dagger }D_{m\omega }=(\lambda +2i\omega r)/\Delta \,. 
\]
Using this equation to reduce the order of differentiation, and recalling 
$D_{m\omega }=D_{m\omega }^{\dagger }+2iK/\Delta $, we obtain

\begin{equation}
(D_{m\omega })^{2}[R_{-1}^{\pm }]=(A_{0}/\Delta )D_{m\omega }[R_{-1}^{\pm
}]+(B_{0}/\Delta )R_{-1}^{\pm }\,,  \label{72}
\end{equation}
where 
\[
A_{0}=2iK\,\,,\,\,B_{0}=\lambda +2i\omega r\,. 
\]
We may rewrite Eqs.\ (\ref{71},\ref{72}) as 
\begin{equation}
R_{+1}^{\pm }=-(C^{\pm }/\Delta )(\bar{A}\frac{d}{dr}R_{-1}^{\pm }+\bar{B}
R_{-1}^{\pm })\,,  \label{73}
\end{equation}
where 
\begin{equation}
\bar{A}=A_{0}=2iK\,
\end{equation}
and 
\[
\bar{B}=B_{0}+(iK/\Delta )A_{0}=\lambda +2i\omega r-2K^{2}/\Delta \,\,. 
\]
We now calculate the determinant $W[R_{+1}^{-},R_{+1}^{+}]$, using 
\[
W[R_{+1}^{-},R_{+1}^{+}]=R_{+1}^{-}D_{m\omega
}[R_{+1}^{+}]-R_{+1}^{+}D_{m\omega }[R_{+1}^{-}]\,, 
\]
along with Eqs.\ (\ref{71},\ref{72}). The calculation yields 
\[
W[R_{+1}^{-},R_{+1}^{+}]=C^{+}C^{-}p\Delta ^{-2}W[R_{-1}^{-},R_{-1}^{+}]\,. 
\]
Substituting this and Eq.\ (\ref{73}) into Eq.\ (\ref{AEM}) we obtain 
\begin{eqnarray*}
a^{\pm }(r) &\equiv &C^{\pm }A^{\pm }(r) \\
&=&-(pW[R_{-1}^{-},R_{-1}^{+}])^{-1}(\bar{A}\frac{d}{dr}R_{-1}^{\mp }+\bar{B}
R_{-1}^{\mp })\,.
\end{eqnarray*}
Finally, substituting this in the above equations for $B_{i}^{-}(r^{\prime
}) $ and $H^{\pm }(r,r^{\prime })$ (with the substitution $r\to r^{\prime }$),
we obtain the expressions for these quantities as specified in section 
\ref{sec9}.


\begin{references}
\bibitem{wald1}  It should be pointed out that Wald's derivation in 
Ref.\ \cite{wald} is applicable even on a local basis (R. M. Wald, private 
communication). This implies that, since $\Psi ^{+}$ 
satisfies all the required equations everywhere in 
$\hat{\Sigma }^{+}$, $h_{\alpha \beta }^{+}\equiv \Pi\left[ \Psi ^{+}\right] $
is a valid MP solution throughout this domain. We 
preferred, however, to demonstrate here the validity of $h_{\alpha \beta }^{+}$ 
without referring to these local considerations.
\end{references}
\end{document}